\documentclass[11pt]{article}

\usepackage{graphicx}
\usepackage{color}
\usepackage{amsmath}
\usepackage{amsthm}
\usepackage{amssymb}
\usepackage{amscd}
\usepackage{bbm}

\newcommand{\R}{\mathbb{R}}

\newcommand{\inr}[1]{\bigl< #1 \bigr>}

\newcommand{\E}{\mathbb{E}}

\newcommand{\eps}{\varepsilon}

\usepackage[textheight=610pt,
  textwidth=450pt,
  centering]{geometry}

\newtheorem{thm}{Theorem}[section]
\newtheorem{Definition}[thm]{Definition}
\newtheorem{Corollary}[thm]{Corollary}
\newtheorem{Lemma}[thm]{Lemma}
\newtheorem{Remark}[thm]{Remark}

\title{Improved bounds for sparse recovery from subsampled random convolutions}
\author{Shahar Mendelson\thanks{Department of Mathematics, Technion, Haifa, 32000, Israel, \rm{shahar@tx.technion.ac.il} \& 
Centre for Mathematics and its Applications, The Australian National University, 
Canberra 0200
Australia}, Holger Rauhut\thanks{Chair for Mathematics C (Analysis),
RWTH Aachen University,
\rm{rauhut@mathc.rwth-aachen.de}},
Rachel Ward\thanks{Mathematics Department, University of Texas at Austin, 2515 Speedway, Austin, Texas, 78712,  \rm{rward@ma.utexas.edu}}}
\title{Improved bounds for sparse recovery from subsampled random convolutions} 

\begin{document}

\maketitle

\begin{abstract}
%
We study the recovery of sparse vectors from subsampled random convolutions via $\ell_1$-minimization.
We consider the setup in which both the subsampling locations as well as the generating vector are chosen at random.  For a subgaussian generator with independent entries, we improve previously known estimates:
if the sparsity $s$ is small enough, i.e.,~$s \lesssim \sqrt{n/\log(n)}$, we show that
$m \gtrsim s \log(en/s)$ measurements are sufficient to
recover $s$-sparse vectors in dimension $n$ with high probability, matching the well-known condition for recovery from standard Gaussian measurements.
If $s$ is larger, then essentially $m \geq s \log^2(s)  \log(\log(s)) \log(n)$ measurements are sufficient, again improving over previous estimates.
{Our results are shown via the so-called robust null space property which is weaker than the standard restricted isometry property.} 
Our method of proof involves a novel combination of small ball estimates with chaining techniques {which should be of independent interest.}
\end{abstract}

%

\section{Introduction}

Compressive sensing \cite{carota06,do06-2,fora13} considers the recovery of (approximately)
sparse vectors from incomplete and possibly perturbed linear measurements via efficient algorithms such as $\ell_1$-minimization.
Provably optimal bounds for the minimal number of required measurements in terms of the sparsity have been shown for Gaussian
and, more generally, subgaussian random matrices \cite{badadewa08,chparewi12,do06-2,dota09,fora13,mepato07,mepato09,ruve08}.

Practical applications demand for structure in the measurement process which is clearly not present in Gaussian random matrices with independent entries. Several types of structured random matrices have been studied,
including random partial Fourier matrices \cite{carota06,cata06,ruve08,ra10,rawa16,bo14},
partial random circulant matrices (subsampled random convolutions) \cite{krmera14,ra09,ra10,rarotr12,ro09-1},
time-frequency structured random matrices \cite{krmera14,pfra10,pfratr11}, and more \cite{aira14,hurast14}.
In this article, we improve known recovery results for partial random circulant matrices.

In mathematical terms, linear measurements of a signal (vector) $x \in \R^n$ can be written as
\[
y = A x \quad \mbox{ with } A \in \R^{m \times n},
\]
and we are particularly interested in the case $m < n$. Compressive sensing predicts that this system can
be solved for $x$ using efficient algorithms if $x$ is sparse enough, say in the sense that $\|x\|_0 = |\{\ell : x_\ell \neq 0\}|$ is small.
While $\ell_0$-minimization is NP-hard \cite{fora13},
several tractable algorithms have been introduced as alternatives, most notably $\ell_1$-minimization (basis pursuit) \cite{carota06,chdosa98,do06-2,fora13} which produces a minimizer of
\[
\min_{z \in \R^n} \|z\|_1 \mbox{ subject to } Az = y.
\]
If $A \in \R^{m \times n}$ is a random draw of a Gaussian matrix, i.e., all entries are standard normal random variables,
then with probability at least $1-e^{-cm}$, any $s$-sparse vector $x \in \R^n$, (i.e., $\|x\|_0 \leq s$), can be reconstructed in a stable way (see below) using $\ell_1$-minimization from the given data $y = Ax$,  provided that
\begin{equation}\label{m:bound}
m \geq C s \ln(en/s)
\end{equation}
for some absolute constant $C > 0$. This bound is optimal \cite{do06-2,foparaul10,fora13} in the sense that the combination of any recovery algorithm
with any measurement matrix requires at least \eqref{m:bound} many measurements in order to achieve stable reconstruction, i.e.,
for any $x \in \R^n$ (not necessarily $s$-sparse), the reconstruction $x^\sharp$ obtained from $y = A x$, satisfies
\begin{equation}\label{stab:reconstruct}
\| x - x^\sharp\|_1 \leq C \sigma_s(x)_1 := C \inf\{ \|x-z\|_1 : \|z\|_0 \leq s\},
\end{equation}
see \cite[Theorem~2.7]{foparaul10} for details. Moreover, \textit{exact} $s$-sparse recovery via $\ell_1$-minimization necessarily requires \eqref{m:bound}, see \cite[Lemma~2.4]{foparaul10}.

Unfortunately, Gaussian random matrices are not suitable for many applications of compressive sensing -- because of their lack of structure.
In fact, structure is required in order to model realistic measurement scenarios and also to speed up the matrix-vector-multiplications that have to be applied many times in known $\ell_1$-minimization algorithms.

An important example of structured random matrices are $m \times n$ matrices that are generated
from the $n \times n$ discrete Fourier (more generally, from a Hadamard type matrix, see Definition~\ref{def:Had-type} below), 
by randomly subsampling $m$ rows. This corresponds to taking $m$ random samples of the discrete Fourier transform of a vector. Again, $\ell_1$-minimization successfully recovers $s$-sparse vectors
with probability at least $1-\varepsilon$ provided that 
\[
m \geq Cs \max\{\log^2(s) \log(n), \log(\varepsilon^{-1})\},
\]
see, for example, \cite{cata06,ruve08,ra10,bo14,hare15}. 

In this article we will be concerned with subsampled random convolutions. The circular convolution on $\R^n$ is defined for two
vectors $x, \xi \in \R^n$ as
\[
(x * \xi)_k = \sum_{j = 1}^n x_j \xi_{k-j \operatorname{mod} n +1}, \quad k=1,\hdots,n.
\]
For a subset $\Omega \subset \{1,\hdots,n\} =:[n]$ of cardinality $m$, let $P_\Omega : \R^n \to \R^m$ be the projection onto the coordinates
indexed by $\Omega$, i.e., $(P_\Omega x)_j = x_j$ for $j \in \Omega$. A subsampled convolution is defined as
\begin{equation}\label{def:B}
B x = 
P_\Omega(x * \xi)
\end{equation}
and the corresponding matrix is a partial circulant matrix generated by $\xi$. Subsampled random convolutions
find applications in radar and coded aperture imaging \cite{bajwa2007toeplitz, haupt2010toeplitz, ro09-1, tropp2006random}, as well as in fast dimensionality reduction maps \cite{krahmer2011new}.

It was shown in \cite{krmera14} that if $\xi$ is a (standard) Gaussian vector
and $\Omega$ is an arbitrary (deterministic) subset of cardinality $m$, then with probability at least $1-\varepsilon$,
every $s$-sparse vector can be reconstructed
from $Bx$ via $\ell_1$-minimization if
\begin{equation}\label{bound:m:circulant}
m \geq C s \max\{\log^2(s) \log^2(n), \log(\varepsilon^{-1})\}.
\end{equation}
{{Stability}} in the sense of \eqref{stab:reconstruct}
holds for such matrices, and the results are robust when the given measurements are corrupted by noise (see more details below). Moreover, the recovery result can be extended to circulant matrices generated by a subgaussian random vector -- an object of central importance to our discussion which will be defined later.

Our focus is on sparse recovery via subsampled random convolutions, where the set $\Omega$ is chosen at random via independent selectors: 
let $(\delta_i)_{i=1}^n$ be independent, $\{0,1\}$-valued random variables with mean
$\delta = m/n \in (0,1]$, and set $\Omega=\{i : \delta_i =1\}$. Then the expected size of $\Omega$ is $\E|\Omega| = m$ and it follows from Bernstein's inequality that $m/2 \leq |\Omega| \leq 3m/2$ with probability
at least $1-2 \exp(-m/9)$.

For the sake of simplicity of this exposition,
we shall first formulate our main theorem for a standard Gaussian generator, i.e., a random vector with independent, mean zero, variance one, normally distributed coordinates. However, the proof we present holds
for more general $L$-subgaussian random vectors with independent coordinates and a more general {class} of random matrices (see Theorem \ref{thm:iid-subgaussian}).

\begin{thm}\label{thm:main} Let $\xi \in \R^n$ be a {draw}
of a standard Gaussian random vector and let $\Omega \subset [n]$ be chosen at random, using independent selectors of mean $\delta = m/n$. Let $B$ be the corresponding partial random circulant matrix defined in \eqref{def:B}.
Let $s \leq c_1 \frac{n}{\log^4(n)}$ and assume that
\begin{equation}\label{bound:m:main}
\begin{array}{ll} m \geq c_3 s \log(en/s) \quad & \mbox{ if } s \leq c_2 \sqrt{\frac{n}{\log(n)}} \\
m \geq c_3 s \log(en/s) \alpha_s^2 \log(e \alpha_s) \quad & \mbox{ if } c_2 \sqrt{\frac{n}{\log(n)}} \leq s \leq c_1 \frac{n}{\log^4(n)},
\end{array}
\end{equation}
where $\alpha_s = \log\left(\frac{s^2}{n} \max\{ \log(en/s), \log(s)\}\right)$.
Then with probability at least
\[
1- 2 \exp\left(-c_0 \min\left\{\frac{n}{s}, s \log(en/s)\right\}\right)
\]
the following holds.
For all $x \in \R^n$, all $e \in \R^m$ with $\|e\|_2 \leq \eta$
and $y = Bx + e$, the minimizer $x^\sharp$
of
\begin{equation}\label{bpdn}
\min \|z\|_1 \quad \mbox{ subject to } \|Bz - y\|_2 \leq \eta
\end{equation}
satisfies
\begin{align}
\|x-x^\sharp\|_1 &\leq C \sigma_s(x)_1 + D \frac{\sqrt{s} \eta}{\sqrt{m}} \qquad {\rm and} \label{err:l1}
\\
\|x-x^\sharp\|_2 & \leq C \frac{\sigma_s(x)_1}{\sqrt{s}} + D \frac{\eta}{\sqrt{m}}. \label{err:l2}
\end{align}
{{The constants $c_1, c_2, c_3, C, D > 0$ are absolute.}}
\end{thm}

Our estimates indicate a phase-transition that occurs when $s$ is roughly of the order of $\sqrt{n}$. Below this level, the partial circulant matrix exhibits the same behavior as the Gaussian matrix (which is the optimal scaling of the number of measurements $m$ as a function of the sparsity parameter $s$) -- it requires $C s \log(en/s)$ measurements
to recover an $s$ sparse vector. Above that level, {our estimates require more measurements}; for example, if $s=n^{\alpha}$ for $1/2 < \alpha < 1$ then $c(L,\alpha) 
{s} \log^3n \cdot \log \log n$ measurements are needed. 

As we will see later, the phase transition at $\sqrt{n/\log(n)}$ is not a coincidence -- the analysis required in the low-sparsity case is truly different from the one needed to deal with the high-sparsity one.
However, it is presently not clear whether the analysis for the high-sparsity case can be improved in order to remove the additional logarithmic factors.

In both cases (low and high sparsity) we improve the estimates from \cite{krmera14}, though it should be noted that \eqref{bound:m:circulant} applies to \textit{any} set $\Omega \subset [n]$ of cardinality $m$, while \eqref{bound:m:main} applies only to randomly chosen $\Omega$. A random selection $\Omega$ has been considered in \cite{ro09-1}, but the estimates there require $m \geq Cs \log^6(n)$. On the other hand, \cite{ro09-1} applies to vectors that
are sparse in an arbitrary (fixed) orthonormal basis and not only in the canonical basis; our proof technique does not seem to extend to this case in a simple way.

{{We}} stress that \eqref{thm:main} provides a uniform recovery guarantee in the sense that a single random draw of the partial circulant matrix is able
to recovery all $s$-sparse vectors simultaneously. This is in contrast to other previous so-called nonuniform results found in the literature \cite{ra09,ra10,jara15} that
only imply recovery of a fixed sparse vector from a random draw of the matrix. Moreover, these nonuniform results give no or weaker stability estimates than \eqref{err:l1} and \eqref{err:l2}, 
see e.g.~\cite[Theorem 4.33]{fora13} or \cite{dora16}.

Another improvement on known estimates is that our results hold for noisy measurements when the noise is bounded in $\ell_q$ for $q \geq 2$ and the $\ell_2$-constraint in \eqref{bpdn} is replaced by an $\ell_q$-constraint (and the error estimates scale with the $\ell_q$ norm of the noise), see Theorem~\ref{thm:iid-subgaussian} 
for details. {{(Note that a standard RIP-based argument appeared after the first submission of this manuscript in \cite[Theorem A.1]{dijura17}}.)} This allows us, for example, to explore quantized compressive sensing (see, e.g., \cite{dilera15}), when the quantization error has a natural $\ell_\infty$-bound. 

We note that for a few other constructions of structured random matrices (with fast matrix-vector multiplication), recovery results with the optimal number of measurements \eqref{m:bound} have been shown under similar size restrictions on the sparsity
as in our main theorem above \cite{aili09,aira14}.
However, it seems that our construction is the simplest one and is arguably the only one among these which models a physically realizable measurement device.
In contrast to these previous results, we are able to extend our bounds to the near-linear sparsity regime at the cost of some additional logarithmic factors.

{{Apart from}} our main results themselves, we believe that our proof techniques {are of independent interest}.
In fact, the crucial ingredient is a probabilistic lower bound on terms of the form {$\inf_{v \in V_r} \|\Gamma_v \xi\|_2$}, where $\Gamma_v$ are matrices indexed by a set of unit-norm $r$-sparse vectors, and $\xi$ is a subgaussian random vector with independent coordinates. It is based on a new approach that can be generalized to other generating random vectors with heavier tails and less independence assumptions, for instance, to log-concave random vectors. 


The article is structured as follows. Section~\ref{sec:prelim} discusses preliminaries such as the null space property, subgaussian random vectors,
states the main result and gives a brief explanation of its proof. Section~\ref{sec:smallball} introduces the small ball estimates required for the proof
as well as moment estimates for norms of subgaussian random vectors. It further provides some covering number estimates required in the sequel.
Section~\ref{sec:main-technical} provides the main technical ingredient of the proof of our main results{{:}} 
{a lower bound} for $\inf_{v \in V_r} \|\Gamma_v \xi\|_2$. Section~\ref{sec:onesparse} provides an upper bound {for} one-sparse vectors, which is {the final ingredient in} the proof of 
the null space property, see also Theorem~\ref{thm:LM14}.

\section{Preliminaries and {the} main result}
\label{sec:prelim}

\subsection{The null space property}
Our analysis is based on a robust version of the null space property{{,}} which is a sufficient and necessary condition for sparse recovery via $\ell_1$-minimization. This version is stable when passing to approximately sparse vectors
{and robust} when the measurements are noisy. 

Given $v \in \R^n$ and $S \subset [n] = \{1,\hdots,n\}$,  let $v_S \in \R^n$ with entries $(v_S)_j = v_j$ for $j \in S$ and $(v_S)_j = 0$ for $j \notin S$. Further, $S^c = [n] \setminus S$ denotes the complement of $S$.

\begin{Definition}
For $1\leq q \leq \infty$, a matrix $A$ satisfies the $\ell_q$-robust null-space property of order $s$ with constants $\nu \in (0,1)$ and
$\tau > 0$ if
$$
\| v_S \|_2 \leq \frac{\nu}{\sqrt{s}} \|v_{S^c}\|_1 + \tau \| A v \|_q
$$
for every $v \in \R^n$ and every $S \subset [n]$ 
of cardinality at most $s$.

\end{Definition}

The following result is standard by now (see, e.g., \cite[Theorem 4.22]{fora13}). It uses the notion of the error of best $s$-term approximation, defined as
\[
\sigma_s(x)_1 = \min_{z : \|z\|_0 \leq s} \|x-z\|_1;
\]
that is, $\sigma_s(x)_1$ is the $\ell_1$-distance between $x$ and the set of $s$-sparse vectors.

\begin{thm}\label{thm:nsp}
Let $1 \leq q \leq \infty$ and let $A$ satisfy the $\ell_q$-robust null space property of order $s$ with constants $\nu \in (0,1)$ and $\tau > 0$. Let $\|e\|_q \leq \eta$, $x \in \R^n$ and put $y = Ax + e$. Then a minimizer $x^\sharp$ of
\[
\min_{z \in \R^n} \|z\|_1 \quad \mbox{ subject to } \|A z - y \|_q \leq \eta
\]
satisfies
\begin{align}
\label{l1:err:bound}\| x - x^\sharp\|_1 & \leq C \sigma_s(x)_1 + D \sqrt{s} \eta\\
\label{l2:err:bound} \| x-x^\sharp\|_2 & \leq \frac{C}{\sqrt{s}} \sigma_s(x)_1 + D \eta,
\end{align}
where the constants are given by $C = \frac{(1+\nu)^2}{1-\nu}$ and $D = \frac{(3+\nu)}{1-\nu} \tau$.
\end{thm}

Roughly speaking, even if $x$ is not $s$-sparse, but only approximated by an $s$-sparse vector, and
if one receives linear measurements of $x$ (i.e., $Ax$) that are corrupted by the ``noise" $e$, then a solution to the minimization problem still yields a good approximation of $x$ if $A$ possesses
the null space property. 
In particular, if $x$ is $s$-sparse then  $\sigma_s(x)_1 = 0$, and if $\eta = 0$ (no noise), then the reconstruction
via equality constrained $\ell_1$-minimization is exact.

In order to show the $\ell_q$-robust null-space property, we will proceed in the following way. Let
\[
{\mathcal{T}}_{\nu,s} := \left\{ v \in \R^n : \|v_S\|_2 \geq \frac{\nu}{\sqrt{s}}\|v_{S^c}\|_1 \right\}.
\]
One may show (see, e.g., \cite{dilera15,kara15}) that if
\[
\inf_{x \in \mathcal{T}_{\nu,s} \cap S^{n-1}} \|A x\|_q \geq \frac{1}{\tau},
\]
then $A$ satisfies the $\ell_q$-robust null space property with constants $\nu$ and $\tau$. Moreover, if we set
$$
V_s= \{x \in \R^n : \|x\|_0 \leq s, \|x\|_2 = 1\}
$$
to be the set of $s$-sparse vectors in the unit sphere, then (\cite[Lemma 3]{kara15}, see also \cite{ruve08})
\begin{equation}\label{embed:NSP}
\mathcal{T}_{\nu,s} \cap S^{n-1} \subset (2+ \nu^{-1}) \operatorname{conv} V_s,
\end{equation}
allowing one to study $\operatorname{conv} V_s$ instead of $\mathcal{T}_{\nu,s}$, where
$\operatorname{conv} S$ denotes the convex hull of {a} set $S$, that is, the set of all convex combinations of finite subsets of $S$.

It turns out that one may replace $\operatorname{conv} V_s$ with $V_r$ for $r$ sufficiently large by adding a condition on one-sparse vectors. This was observed for $q=2$ in \cite[Lemma 5.1]{im13} (see also \cite[Theorem B]{leme14}), {but extends also to $q > 2$ as we outline below}.

\begin{thm} \label{thm:LM14}
Let $A \in \R^{m \times n}$ satisfy
\begin{equation} \label{eq:tau-and-M}
\inf_{x \in V_{r}} \|A x\|_2 \geq \tau^{-1} \quad {\rm and} \quad  \max_{j \leq [n]} \|A e_j \|_2 \leq M.
\end{equation}
If $c(\nu)=\nu^2/(2\nu+1)^2$ and
$$
s \leq c(\nu) \frac{r-1}{M^2\tau^2-1},
$$
then
$$
\inf_{x \in \mathcal{T}_{\nu,s}} \|Ax\|_2 \geq \frac{1}{\sqrt{2} \, \tau}.
$$
{Consequently, $A$ also satisfies the $\ell_q$-robust NSP, for $q > 2$:
$$
\inf_{x \in \mathcal{T}_{\nu,s}} \|Ax\|_q \geq \frac{1}{ m^{1/2 - 1/q}} {\inf_{x  \in \mathcal{T}_{\nu,s}}} \| Ax \|_2 \geq \frac{1}{m^{1/2 - 1/q} \sqrt{2} \tau}.
$$
}
\end{thm}
\proof
By \eqref{embed:NSP}, it suffices to show that the conditions in \eqref{eq:tau-and-M} imply that
\[
\inf_{x \in (2+ \nu^{-1}) \operatorname{conv} V_s \cap S^{n-1}} \|A x\|_2 \geq 1/(\sqrt{2} \tau).
\]
Applying \cite[Lemma 2.6]{leme14}, 
it follows from \eqref{eq:tau-and-M} that for any $y \in \R^n$
\begin{equation}\label{nsq-lp-H}
\|A y\|_2^2  \geq \tau^{-2} \|y\|_2^2 - \frac{1}{r-1}\left(\|y\|_1 \sum_{j=1}^n \|A e_j\|_2^2 |y_j| - \tau^{-2} \|y\|_1^2\right).
\end{equation}
Let $B_1^n$ be the unit ball in $\ell_1^n$ and observe that $\operatorname{conv} V_s \subset \sqrt{s} B_1^n$. Thus, if $c_1(\nu)=2+1/\nu$ and $y \in S^{n-1} \cap c_1(\nu) \operatorname{conv} V_s$,
$$
\|y\|_2=1 \ \ \ {\rm and} \ \ \ \|y\|_1 \leq c_1(\nu)\sqrt{s}.
$$
Therefore,
\begin{align}
\|A y \|_2^2 &\geq \tau^{-2}\left(1- \frac{\|y\|_1^2}{r-1}(\tau^2 \max_{1 \leq j \leq n} \|Ae_j\|_2 - 1)\right) \nonumber \\
&\geq \tau^{-2}\left(1- \frac{c_1^2(\nu)s}{r-1} \left(\tau^2 M^2-1\right)\right) \geq \frac{1}{2\tau^2} \nonumber
\end{align}
by our choice of $s$. {The first inequality in \eqref{nsq-lp-H} is due to H{\"o}lder.}
\endproof

With Theorem \ref{thm:LM14} at hand, we will take the following course of action: we will 
show that
\begin{equation} \label{eq:condition-on-A}
\inf_{x \in V_{r}} \|P_\Omega A x\|_2 \gtrsim \sqrt{m} \ \ \ {\rm and} \ \ \ \max_{j \in [n]} \|P_\Omega A e_j \|_2 \lesssim \sqrt{m}
\end{equation}
for a partial circulant matrix $P_\Omega A$, whose rows are chosen using iid selectors.

\subsection{Subgaussian random vectors}

Just as in \cite{krmera14} we will focus on generators $\xi$ that are isotropic, $L$-subgaussian and have independent coordinates.
\begin{Definition} \label{def:L-subgaussian}
A centered random vector $\xi=(\xi_i)_{i=1}^n$ is $L$-subgaussian if, for every $x \in \R^n$,
$$
{(\E |\inr{\xi,x} |^{p} )^{1/p} \leq L \sqrt{p} (\E |\inr{\xi,x} |^{2} )^{1/2}}.
$$
\end{Definition}
Assume that $\xi_1,\hdots,\xi_n$ are independent, mean-zero, variance $1$, $L$-subgaussian random variables. In other words, for every $p \geq 1$,
$$
{( \E | \xi_i |^{p} )^{1/p} \leq L \sqrt{p} (\E | \xi_i |^{2} )^{1/2} = L \sqrt{p}}.
$$
Then $\xi$ is an {\it isotropic} random vector on $\R^n$: for every $x \in \R^n$, $\E \inr{\xi,x}^2 =\|x\|_2^2$, and it is standard to verify that it is $L$-subgaussian as well (see, e.g., \cite{ve12}). 
Is it straightforward to show that if $\xi$ is an $L$-subgaussian random vector, then for every $x \in \R^n$ and $u \geq 1$,
$$
Pr\left(|\inr{\xi,x}| \geq u L (\E |\inr{\xi,x} |^{2} )^{1/2} \right) \leq 2\exp(-cu^2),
$$
for a suitable absolute constant $c$. If, in addition, $\xi$ is isotropic, then 
$$
Pr\left(|\inr{\xi,x}| \geq u L \|x\|_2 \right) \leq 2\exp(-cu^2).
$$

\subsection{Main result}

Our main result provides a bound for the null-space property of partial random circulant matrices.
It implies Theorem~\ref{thm:main} 
via Theorem~\ref{thm:nsp}.

Recall that
\begin{description}
\item{$\bullet$} $\xi$ is a random vector whose coordinates are independent, mean-zero, variance $1$, $L$-subgaussian random variables;
\item{$\bullet$} $A$ is the complete circulant matrix generated by $\xi$;
\item{$\bullet$} $(\delta_i)_{i=1}^n$ are independent selectors of mean $\delta=m/n$, and $\Omega=\{i :\delta_i=1\}$.
\end{description}

\begin{thm} \label{thm:iid-subgaussian}
There exist constants $c_0,\hdots,c_6$ that depend only on $L$, $\rho$ and $\tau$ for which the following holds {for all $q \geq 2$}. With probability at least
$$
1-2\exp\left(-c_0 \min\left\{\frac{n}{r},r\log\left(\frac{en}{r}\right)\right\}\right),
$$
the partial circulant matrix $P_\Omega A$ 
satisfies the {$\ell_q$}-robust null-space property of order $r$ with constants $\rho \in (0,1)$ and {$\frac{\tau}{m^{1/q}}$}, where
$$
\delta n =c_3 r \log\left(\frac{en}{r}\right) \ \ \ {\rm if} \ \ r \leq c_4 \sqrt{\frac{n}{\log n}},
$$
and
$$
\delta n = c_3 r \log\left(\frac{en}{r}\right) \cdot \alpha_r^2 \log(e \alpha_r) \ \  {\rm if} \ \   c_4\sqrt{\frac{n}{\log n}} < r \leq c_5 \frac{n}{\log^4 n}.
$$
The constant $\alpha_r$ satisfies
$$
\alpha_r \leq \log \left(c_6 \frac{r^2}{n} \max \left\{\log\left(\frac{en}{r}\right), \log (er) \right\} \right).
$$
\end{thm}
{Theorem \ref{thm:iid-subgaussian} is obtained from the analysis to follow by combining Theorem \ref{thm:proj-gamma-e-i} and Corollary \ref{cor:from-structure-to-proj}, and applying Theorem \ref{thm:LM14}. 
Actually, our analysis holds for a more general class of random matrices, see Definition~\eqref{def:Had-type} and the remarks following it.  
}

\subsection{The heart of the argument}
The proof of Theorem \ref{thm:iid-subgaussian} has two main components. We will begin by analyzing the way a complete circulant matrix $A:\R^n \to \R^n$ generated by $\xi$ acts on $V_r$, and then apply a random ``selector projection" $P_\Omega$ to the image $A V_r$. Our primary goal is to obtain a lower bound on
\begin{equation} \label{eq:cond-inf}
\inf_{v \in V_r} \|P_\Omega A v\|_2^2 = \inf_{v \in V_r}\sum_{i=1}^n \delta_i \inr{A v , e_i}^2.
\end{equation}

Thanks to the nature of circulant matrices, there is a standard representation of $\{Av : v \in V_r\}$ via the Fourier transform.
Let ${\cal F}$ be the (un-normalized) 
Fourier transform, i.e., ${\cal F}_{j,k} = e^{-2\pi i jk/n}$, (which we treat as a ``real operator'' from $\R^n$ onto the image of $\R^n$ in order to avoid working with $\mathbb{C}^n$)  
and set $\hat{v} = {\cal F}v$. If $D_x={\rm diag}(x_1,\hdots,x_n)$ then
$$
Av={\cal F}^{-1} D_{\hat{v}} {\cal F} \xi.
$$
We will consider a more general set of matrices:
\begin{Definition} \label{def:Had-type}
An orthogonal matrix $O$ is of Hadamard type with constant $\beta \geq 1$ if for every $i,j \in [n]$, $|O_{i,j}| \leq \beta/\sqrt{n}$.
\end{Definition}
In what follows we will fix three matrices, $U,W$ and $O$, each of which is of Hadamard type with constant $\beta$, and for $x \in \R^n$ we set
\begin{equation}\label{def:gen:matrix}
\Gamma_x = \sqrt{n} UD_{Wx} O,
\end{equation}
where $D_{Wx} = {\rm diag} ((\inr{W_i,x})_{i=1}^n)$.
Clearly, the representation of $Av$ is precisely of this form: if $A$ is the complete circulant matrix with the generator $\xi$ then for every $v \in \R^n$, $Av = \sqrt{n} UD_{Wv}O\xi$ for the choice of $U=n^{-1/2}{\cal F}^{-1}$ and $W=O=n^{-1/2}{\cal F}$; in this case $\beta=1$.

From here on, for $V \subset \R^n$ set
$$
\Gamma_V= \{\sqrt{n}UD_{Wv}O : v \in V\};
$$
naturally, the set of matrices we will be interested in is $\Gamma_{V_r}$.

Observe that if $\xi$ is an isotropic random vector  then for every $v \in S^{n-1}$, $\E \|\Gamma_v \xi\|_2^2 = n$, and at least on average, for a single vector $v \in V_r$, one expects to have $\|\Gamma_v \xi\|_2 \sim \sqrt{n}$.

Unfortunately, showing that $\inf_{v \in V_r} \|\Gamma_v \xi\|_2 \geq c\sqrt{n}$ {alone} does not lead to a nontrivial lower bound on \eqref{eq:cond-inf}. To see why, set
$$
x^1=(\sqrt{n},0,\hdots,0) \ \ {\rm and} \ \ x^2=(1,\hdots,1).
$$
Both $x^1$ and $x^2$ have a Euclidean norm of $\sqrt{n}$, but any attempt of selecting a random subset of coordinates of cardinality $m \ll n$ fails miserably for $x^1$ and succeeds for $x^2$: typically, $P_\Omega x^1=0$ while $\|P_\Omega x^2\|_2 =\sqrt{m}$. We will be looking at this type of ``good behavior", exhibiting a (one sided) {\it  standard shrinking} phenomenon. The term ``one-sided standard shrinking" used in this context usually refers to a random projection operator $T$ of rank $m$, for which, with high probability,
$$
\|T v\|_{2} \geq c\sqrt{m/n} \|v\|_{2} \quad \mbox{ for all vectors of interest}.
$$
The operator we are interested in is indeed random, and of the form $T=P_{\Omega}$ -- a random coordinate projection -- but as the example of $x^1$ shows, $P_{\Omega}$ may map $x$ to $0$ even if $x$ has a large norm -- unless one imposes some additional condition on $x$.

The condition we will focus on here is that $x$ has a regular coordinate structure, that is, for suitable constants $\alpha$ and $\theta$,
\begin{equation} \label{eq:reg-vector}
\left|\left\{i :  |x_i| \geq \|x\|_2\frac{\alpha}{\sqrt{n}} \right\} \right| \geq \theta n.
\end{equation}
The notion of regularity in \eqref{eq:reg-vector} implies that $|x_i|$ is at least $\sim \|x\|_2/\sqrt{n}$ for a large subset of coordinates -- of cardinality that is proportional to the dimension $n$.  That set of coordinates contributes at least a proportion of the Euclidean norm of $x$, and moreover, for a random choice of $\Omega \subset [n]$, $\|P_\Omega x\|_2 \geq c\|x\|_2 \cdot \sqrt{|\Omega|/n}$ with high probability.

Thus, in addition to showing that $\inf_{v \in V_r} \|\Gamma_v \xi\|_2 \geq c\sqrt{n}$, we will prove that each of the vectors $\Gamma_v \xi$ is regular in the sense of \eqref{eq:reg-vector}. We will do so by representing a typical realization of the set $\{\Gamma_v \xi : v \in V_r\}$ as a subset of the Minkowski sum of two (random) sets $T_1+T_2$ 
defined in the following way: Let $H \subset V_r$ be a fine enough net with respect to the Euclidean distance and set $T_1=\{\Gamma_x \xi : x \in H\}$. For each $v \in V_r$, choose $x = \pi(v) \in H$ minimizing
$\|x-v\|_2$. Then 
$T_2=\{\Gamma_{v-\pi(v)} \xi : v \in V_r\}$.
We will show the following properties of the sets $T_1,T_2$: 
\begin{description}
 \item{$\bullet$} Every $t \in T_1$ satisfies $\|t\|_2 \gtrsim \sqrt{n}$ and has regular coordinate structure in the sense of \eqref{eq:reg-vector}. 
 As a consequence, a random coordinate projection $P_\Omega$ will not shrink the $\ell_2$-norm of elements in $T_1$ by more than a factor of $\sim \sqrt{\delta}$, and
 therefore, with high probability,
    $$
    \inf_{x \in H} \|P_\Omega \Gamma_x \xi\|_2 \geq c\sqrt{\delta n}.
    $$
\item{$\bullet$} The set of ``random oscillations" $T_2$ has Euclidean diameter smaller than $(c/2)\sqrt{\delta n}$. Thus, its effect is negligible.
\end{description}


What may still appear mysterious is the claim that there is a phase transition in the choice of $\delta$ -- and thus in the required number of measurements. The origin of the phase transition lies in a gap between the cardinality of the net $H$ and the probability estimate one is likely to have for each $\Gamma_v$. Indeed, for reasons that will be clarified later, the probability that $\Gamma_v \xi$ is ``well behaved" can be estimated by $\exp(-cn/r)$. In contrast, as a nontrivial Euclidean net in $V_r$, $|H| \geq \exp(c_1r\log(en/r))$. In the low-sparsity case, when $n/r \gtrsim r\log(en/r)$, the individual probability estimate is strong enough to allow uniform control on all the vectors in the net $H$. In the high-sparsity case that is no longer true, and an additional argument is required to bridge the gap between $n/r$ and $r\log(en/r)$. Specifically, we will show how one may ``transfer information" from a set of cardinality $\exp(cn/r)$ to the net $H$ whose cardinality is much larger -- of the order of $\exp(cr\log(en/r))$.

\subsection{Notation}

Throughout this article, absolute constants are denoted by $c,c_1,C$, etc. 
The notation
$c(L)$ refers to a constant that depends only on the parameter $L$; $a \sim b$ implies that there are absolute constants $c$ and $C$ for which $ca \leq b \leq Ca$; and $a \sim_L b$ means that the constants $c$ and $C$ depend only on $L$. The analogous one-sided notation is $a \lesssim b$ and $a \lesssim_L b$. Constants whose values remain unchanged throughout the article are denoted by $\kappa_1$, $\kappa_2$, etc. .

For $1 \leq p \leq \infty$ let $\ell_p^n$ be the normed space $(\R^n,\| \ \|_p)$ and set $B_p^n$ to be its unit ball. $S^{n-1}$ is the Euclidean unit sphere in $\R^n$. The expectation is denoted by $\E$ and $Pr$ denotes the probability of an event. The $L_p$-norm of a random variable $X$ is denoted $\|X\|_{L_p} = (\E |X|^p)^{1/p}$. We also recall that $[n] = \{1,\hdots,n\}$.

\section{Small ball estimates and chaining}
\label{sec:smallball}

\subsection{The random generator}
Recall that the random vector $\xi$ we are interested in has independent coordinates $(\xi_i)_{i=1}^n$ that are mean-zero, variance $1$ and $L$-subgaussian. In particular, $\xi$ is an isotropic, $L$-subgaussian random vector.

A simple observation is that the $\xi_i$'s satisfy a small-ball property: there are positive constants $c_1$ and $c_2$ that depend only on $L$ for which
\begin{equation} \label{eq:small-ball-xi-i}
\sup_{u \in \R} Pr(|\xi_i-u| \geq c_1) \geq c_2.
\end{equation}
Indeed, for any $u \in \R$, $\|\xi_i-u\|_{L_2} \sim \max\{\|\xi_i\|_{L_2},|u|\}$, and thus $\|\xi_i-u\|_{L_4} \leq c_3L\|\xi_i-u\|_{L_2}$ for a suitable absolute constant $c_3$. The small-ball property \eqref{eq:small-ball-xi-i} is an immediate outcome of the Paley-Zygmund inequality (see, e.g., \cite[Lemma 7.16]{fora13})  applied to each $X_u=|\xi_i-u|$.

The small-ball property \eqref{eq:small-ball-xi-i} tensorizes, leading to a vector small-ball property for $\xi=(\xi_i)_{i=1}^n$. To formulate this property, let $\|\Gamma\|_{HS}$ and $\|\Gamma\|_{2 \to 2}$ denote the Hilbert-Schmidt (Frobenius) and operator norms of a matrix $\Gamma$, respectively, and set 
$$
d_{\Gamma} = \left(\frac{\|\Gamma\|_{HS}}{\|\Gamma\|_{2 \to 2}}\right)^2.
$$
\begin{thm} \label{thm:RV-small-ball} \cite{ruve15}
{For $0<p<1$, there exists a constant $c=c(p)>0$ for which the following holds.}
Let $X_1,\hdots,X_n$ be independent random variables that satisfy for some $t>0$ and $0<p<1$
$$
\sup_{u \in \R} Pr(|X_i-u| \leq t) \leq p.
$$
Then, for $\mathbb{X}=(X_1,\hdots,X_n)$ and every matrix $\Gamma:\R^n \to \R^m$,
$$
Pr( \|\Gamma \mathbb{X}\|_2 \leq t\|\Gamma\|_{HS}) \leq \left(\frac{1}{2}\right)^{c d_\Gamma}.
$$
\end{thm}
The small-ball property for individual $\xi_i$'s from \eqref{eq:small-ball-xi-i} and Theorem \ref{thm:RV-small-ball} imply that the random vector $\xi$ satisfies a small-ball estimate.
\begin{Corollary} \label{cor:xi-small-ball}
There exist constants $\kappa_1$ and $\kappa_2$ that depend only on $L$ such that, for any matrix $\Gamma:\R^n \to \R^m$,
\begin{equation} \label{eq:small-ball-xi}
Pr(\|\Gamma \xi\|_2 \leq \kappa_1 \|\Gamma\|_{HS}) \leq \left(\frac{1}{2}\right)^{\kappa_2 d_{\Gamma}}.
\end{equation}
\end{Corollary}
It should be noted that a subgaussian vector with independent coordinates is not the only random vector that satisfies a small-ball estimate like \eqref{eq:small-ball-xi}. Moreover, such small-ball estimates can be used to extend our main result to a larger class of generators -- a direction we will not explore further in this work.

The other type of bound we require deals with the way the moments of $\|\xi\|$ grow, for an arbitrary norm $\| \ \|$ on $\R^n$. Unlike Corollary \ref{cor:xi-small-ball}, this feature does not require $\xi$ to have independent coordinates, and it holds for any (isotropic) subgaussian random vector (see, for example, \cite[Theorem 2.3]{krmera14}).
\begin{thm} \label{thm:subgaussian-norm-moments} 
There exists an absolute constant $c$ for which the following holds.
Let $\xi$ be an isotropic, $L$-subgaussian random vector in $\R^n$ and set $G=(g_1,\hdots,g_n)$ to be the standard {G}aussian vector in $\R^n$. Let $\| \ \|$ be a norm on $\R^n$ and set $B^\circ$ to be the unit ball of its dual norm. Then for every $p \geq 1$,
$$
(\E\|\xi\|^p)^{1/p} \leq cL\bigl(\E\|G\| + \sqrt{p} \sup_{t \in B^\circ} \|t\|_2\bigr).
$$
\end{thm}

We will consider two families of norms associated with the non-increasing rearrangement of the coordinates of a vector.
\begin{Definition}
Let $(x_i^*)_{i=1}^n$ denote the non-increasing rearrangement of $(|x_i|)_{i=1}^n$. For $k \in [n]$, set
$$
\|x\|_{[k]} = \max_{|I|=k} \bigl(\sum_{i \in I} x_i^2 \bigr)^{1/2} = \bigl(\sum_{i=1}^k (x_i^*)^2\bigr)^{1/2}.
$$
\end{Definition}
When $k=n$, the norm $\| \ \|_{[k]}$ is simply the Euclidean norm, and the unit ball of the dual norm is just the standard Euclidean unit ball. When $1 \leq k<n$, the dual unit ball consists of the set of unit-$\ell_2$-norm $k$-sparse vectors, that is,
$$
V_k = \{ v \in S^{n-1} : \|v\|_0 \leq k \}.
$$
{In order to} apply Theorem~\ref{thm:subgaussian-norm-moments} to $\| \ \|_{[k]}$ one has to control
$\E\|G\|_{[k]} = \E\bigl(\sum_{i =1}^k (g_i^*)^2 \bigr)^{1/2}$.
The following result for subgaussian random variables does not require independence.
\begin{Lemma} \label{lemma:Klartag} \cite{klartag02}
Let $Z_1,\hdots,Z_n$ be mean-zero and $L$-subgaussian random variables such that $\max_{i \in [n]} \|Z_i\|_{L_2} \leq M$. Then
$$
\E \bigl(\sum_{i=1}^k (Z_i^*)^2 \bigr)^{1/2} \leq cLM \sqrt{k \log(en/k)}.
$$
\end{Lemma}
\proof Since the proof of this result is not provided in \cite{klartag02}, we give it here for convenience. Since the $Z_i$ are $L$-subgaussian, there exist constants $c_0, c_1 > 0$ such that (see e.g.~\cite[Proposition 7.23]{fora13}) 
\[
\E \exp\left(c_0 \frac{Z_i^2}{L^2\|Z_i\|_{L_2}^2}\right) \leq c_1.
\] 
By Jensen's inequality and concavity of the logarithm
\begin{align*}
&\left[\E \bigr( \frac{1}{k} \sum_{i=1}^k (Z_i^*)^2 \bigr)^{1/2} \right]^2 \leq \E \frac{1}{k} \sum_{i=1}^k (Z_i^*)^2 \nonumber \\
&\leq c_0^{-1} M^2 L^2 \E \frac{1}{k} \sum_{i=1}^k \log(\exp(c_0 (Z_i^*)^2/(L\|Z_i^*\|_{L_2})^2)) \nonumber \\
&\leq c_0^{-1} M^2 L^2\log \left( \frac{1}{k} \sum_{i=1}^k \E \exp\left( c_0 \frac{(Z_i^*)^2}{L^2\|Z_i^*\|_{L_2}^2}\right) \right) \nonumber \\
&\leq c_0^{-1} M^2 L^2  \log \left( \frac{1}{k} \sum_{i=1}^n \E \exp\left( c_0 \frac{Z_i^2}{L^2\|Z_i\|_{L_2}^2}\right) \right) \leq  c_0^{-1} M^2 L^2 \log( c_1 n/k).
\end{align*}
Rearranging this inequality and adjusting constants yields the claim.
\endproof

For the choice $Z_i=\xi_i$, it follows from Lemma \ref{lemma:Klartag} and Theorem \ref{thm:subgaussian-norm-moments} that for every $p \geq 1$ and $k \in [n]$,
\begin{equation} \label{eq:k-norm-moment}
(\E\|\xi\|_{[k]}^p)^{1/p} \leq cL \left(\sqrt{k \log(en/k)}+ \sqrt{p}\right).
\end{equation}
The second family of norms we require is a generalization of the first one. Let $\Gamma$ be a matrix and set
$$
{\|x\|_{\Gamma,k}}=\|\Gamma x\|_{[k]} = \bigl(\sum_{i=1}^k (\inr{\Gamma x,e_i}^*)^2\bigr)^{1/2}= \sup_{t \in \Gamma ^* V_k} \inr{x,t},
$$
where the last equality follows from $\|z\|_{[k]} = \sup_{t \in V_k} \inr{z,t}$. 
\begin{Lemma} \label{lemma:Gamma-k-norm-moment}
Let $\xi$ be an isotropic, $L$-sugaussian random vector. Then for every matrix $\Gamma$ and any $p \geq 1$,
$$
(\E\|\Gamma \xi\|_{[k]}^p)^{1/p} \leq cL \left(\sqrt{k\log(en/k)} \max_{1 \leq i \leq n} \|\Gamma^*e_i\|_2 + \sqrt{p} \sup_{t \in V_k} \|\Gamma^* t\|_2\right).
$$
\end{Lemma}
\proof
By Theorem \ref{thm:subgaussian-norm-moments}, it suffices to estimate 
\[
{\E \|G\|_{\Gamma,k}} = \E \sup_{t \in \Gamma^* V_k} \inr{{G},t} = \E \bigl(\sum_{i=1}^k (\inr{G,\Gamma^*e_i}^*)^2 
\bigr)^{1/2}
\] 
for {a} standard {G}aussian vector $G$. This expectation may be controlled using Lemma \ref{lemma:Klartag} for the choice of $Z_i=\inr{G,\Gamma^* e_i}$ and that fact that $M=\max_{i \in [n]} \|Z_i\|_{L_2} = \max_{i \in [n]} \|\Gamma^* e_i\|_2$.
\endproof
The following standard relation between moments and tails is recorded for convenience as it will be used frequently in the sequel. Its proof follows immediately from Markov's inequality. 
\begin{Lemma}\label{lem:moments}
Assume that a random variable $Z$ satisfies
$(\E |Z|^p)^{1/p} \leq A$ for some $p > 0$ and $A > 0$.
Then, for $\alpha > 1$,
\[
Pr(|Z| \geq \alpha A) \leq \alpha^{-p}. 
\]
\end{Lemma}

As noted above, the matrices we will be interested in are of the form 
$$\Gamma_v=\sqrt{n}UD_{Wv}O,
$$ 
where $U$, $W$ and $O$ are Hadamard type matrices with constant $\beta$ and $v \in \R^n$. Thus,
{
$$
\|\Gamma_v\|_{HS} = \sqrt{n}\|v\|_2, \ \ \|\Gamma_v\|_{2 \to 2} = \sqrt{n}\|Wv\|_{\infty} \leq \beta \|v\|_0^{1/2} \| v \|_2,
$$
}
and for every $i \in [n]$,
\begin{equation}\label{bound:Gammav:ei}
\|\Gamma^*_v e_i\|_2 = \sqrt{n} \bigl(\sum_{\ell=1}^n \inr{W_\ell,v}^2 \cdot U_{i,\ell}^2\bigr)^{1/2} \leq \beta \|v\|_2.
\end{equation}
Combining Corollary \ref{cor:xi-small-ball}, Lemma \ref{lemma:Gamma-k-norm-moment} and 
Lemma~\ref{lem:moments},
one has the following:
\begin{Corollary} \label{cor:basic-estimates-Gamma-v}
There exist constants $\kappa_1$, $\kappa_2$ and $\kappa_3$ that depend only on $L$ and $\beta$ for which the following holds. If $v \in V_r$ then
\begin{equation} \label{eq:basic-estimate-Gamma-v-1}
Pr(\|\Gamma_v \xi\|_2 \leq \kappa_1\sqrt{n}\|v\|_2) \leq \left(\frac{1}{2}\right)^{\kappa_2 n/ r},
\end{equation}
and if $v \in \R^n$ and then with probability at least $1-\exp(-p)$,
\begin{equation} \label{eq:basic-estimate-Gamma-v-2}
\|\Gamma_v \xi\|_{[k]} \leq \kappa_3 (\|v\|_2 \sqrt{k\log(en/k)} + \sqrt{p} \cdot \sqrt{n}\|Wv\|_{\infty}).
\end{equation}
\end{Corollary}
\begin{Remark}
In what follows, \eqref{eq:basic-estimate-Gamma-v-1} and \eqref{eq:basic-estimate-Gamma-v-2} are the key features of $\xi$ that we will use. To establish those two facts 
we used rather special properties of $\xi$, but while those special properties (for example, that $\xi$ is stochastically dominated by a {G}aussian vector) are highly restrictive, 
\eqref{eq:basic-estimate-Gamma-v-1} and \eqref{eq:basic-estimate-Gamma-v-2}, or even further relaxations of the two, actually hold for a wider variety of random vectors. 
We will pursue this direction in a future contribution.
\end{Remark}

\subsection{Definitions and basic facts} \label{sec:cover}
Let $(T,d)$ be a metric space. A subset $T^\prime \subset T$ is called $\eps$-separated if for every $x,y \in T^\prime$, $d(x,y) \geq \eps$. By a standard comparison of packing and covering numbers, see e.g.~\cite[Lemma C.2]{fora13}, if $T^\prime$ is a maximal $\eps$-separated subset of $T$, it is also an $\eps$-cover: that is, for every $x \in T$ there is some $y \in T^\prime$ for which $d(x,y) \leq \eps$.

In what follows we denote by $N(T,d,\eps)$ the cardinality of a minimal $\eps$-cover of $T$. Note that if $T \subset \R^n$ and $d$ is a norm on $\R^n$ whose unit ball is $B$, then $N(T,d,\eps)$ is the minimal number of translates of $\eps B$ needed to cover $T$. Therefore, we will sometimes abuse notation and write $N(T,\eps B)$ instead of $N(T,d,\eps)$.

We will also use the language of {\it Generic Chaining} \cite{ta14-1} extensively.
\begin{Definition} \label{def:adm-seq}
Given a metric space $(T,d)$, an admissible sequence $(T_s)_{s \geq 0}$ is a sequence of subsets of $T$, with $|T_0|=1$ and $|T_s| \leq 2^{2^s}$. Together with an admissible sequence one defines a collection of maps $\pi_s:T \to T_s$. Usually, $\pi_s t $ is chosen as a nearest point to $t$ in $T_s$ with respect to the metric $d$. For $s \geq 0$ set
$$
\Delta_s t = \pi_{s+1}t -\pi_s t.
$$
\end{Definition}

Let us define several parameters that will be used throughout the proof of Theorem \ref{thm:iid-subgaussian}.

\begin{enumerate}
\item For $r \in [n]$ set
\begin{equation}
 \label{eq:rho}
\rho = 10\log_2 e \cdot \max\left\{1, \frac{\log(er)}{\log(en/r)}\right\}.
\end{equation}
\item Using the notation introduced earlier, put
\begin{equation}\label{def:kappa4}
\kappa_4 = \min\left\{\frac{\kappa_2}{2}, \frac{\kappa_1^2}{
{256}\kappa_3^2 L^2 \beta^2}\right\},
\end{equation}
and observe that $\kappa_4$ depends only on $L$ and $\beta$. Moreover, without loss of generality, $\kappa_4 \leq 1$, and $\kappa_3, L, \beta \geq 1$.

\item Set $s_0$ and $s_1$ to satisfy
\begin{equation}
2^{s_0} = \frac{\kappa_4 n}{r} \ \ \ {\rm and} \ \ \ 2^{s_1}=\rho r \log(en/r) 
\label{eq:s}
\end{equation}
and without loss of generality we will assume that $s_0$ and $s_1$ are integers. 
\item Finally, let
\begin{equation}
\label{eq:alpha}
\alpha_r=\max\left\{1,\log\left(\frac{\rho r \log(en/r)}{\kappa_4 (n/r)}\right)\right\}=\max\{1,\log(2^{s_1-s_0})\}.
\end{equation}
\end{enumerate}

A key part in the proof of Theorem \ref{thm:iid-subgaussian} requires a different argument when $2^{s_0} \geq 2^{s_1}$ and when the reverse inequality holds. As we indicated earlier, we will call the former the ``low-sparsity" case, and the latter the ``high-sparsity" case. It is straightforward to verify that in the low-sparsity case ($2^{s_0} \geq 2^{s_1}$), this corresponds to
$$
r \leq c\kappa_4^{1/2} \sqrt{\frac{n}{\log(cn/\kappa_4)}}, \ \ \rho = 10\log_2 e, \ \ {\rm and} \ \ \alpha_r=1,
$$
while in the ``high-sparsity" case,
$$
r \geq c\kappa_4^{1/2} \sqrt{\frac{n}{\log(cn/\kappa_4)}};
$$
if $r \leq \sqrt{n}$ then $\rho = 10\log_2 e$, and otherwise, $\rho=10\log_2 e \cdot \frac{\log(er)}{\log(en/r)}$. Thus,
$$
\alpha_r = \log\left(\frac{cr^2}{\kappa_4 n} \cdot \log\left(\frac{en}{r}\right)\right) \ \ {\rm if } \ \ r \leq \sqrt{n},
$$
and
$$
\alpha_r = \log\left(\frac{cr^2}{\kappa_4 n} \cdot \log(er)\right) \ \ {\rm otherwise}.
$$
Let us again emphasize that $\kappa_1$, $\kappa_2$, $\kappa_3$ and $\kappa_4$ are all constants that depend only on $L$ and $\beta$ -- an observation that will be used throughout this article.

\subsection{Covering of $V_r$}

Let us begin by constructing (a part of) an admissible sequence for $V_r$.
\begin{Lemma} \label{lemma:covering-sphere}
Let $1 \leq r \leq n/2$ and $s_1$ as above. There exists an admissible sequence $(V_{r,s})_{s \geq s_1}$ for which
$$
\sup_{v \in V_r} \sum_{s \geq s_1} (\sqrt{n} + \sqrt{r}2^{s/2})\|\Delta_s v\|_2 \leq \frac{c}{n^{3/2}},
$$
where $\pi_s v$ is the nearest point to $v$ in $V_{r,s}$ with respect to the Euclidean norm, $\Delta_s v = \pi_{s+1}v - \pi_s v$ and $c$ is an absolute constant.
\end{Lemma}
We note that the exponent $3/2$ above is rather arbitrary. We could easily replace it by a larger one by adjusting constants.

\proof Let $V_{r,s}$ be a maximal $\eps_s$ separated subset of $V_r$ with respect to the Euclidean norm and of cardinality $2^{2^s}$. Thus it is also an $\eps_s$-cover of $V_r$ and
$$
\|\Delta_s v\|_2 \leq \|\pi_{s+1}v -v \|_2 +\|\pi_s v -v \|_2 \leq 2\eps_s.
$$
To estimate $\eps_s$, observe that by a standard volumetric estimate, see e.g.~\cite[Proposition C.3]{fora13}, and summing
over all $\binom{n}{r}$ possible support subsets of $[n]$ of cardinality $r$, for any $0<\eps<1/2$, the cardinality of a maximal $\eps$-separated subset of $V_r$ is at most
{
$$
\binom{n}{r}  \left(1+\frac{2}{\eps}\right)^r \leq \binom{n}{r} \left(\frac{3}{\eps}\right)^r \leq \left(\frac{3en}{r \eps}\right)^r.
$$
}
Hence,
$$
\eps_s \leq 2^{-2^s/r} \left(\frac{3en}{r}\right),
$$
and
\begin{equation*}
\sup_{v \in V_r} \sum_{s \geq s_1} 2^{s/2} \|\Delta_s v\|_2 \leq 2\sum_{s \geq s_1} 2^{s/2} \eps_s \leq c_0 \left(\frac{n}{r}\right) \cdot \sum_{s \geq s_1} 2^{s/2-2^s/r}. 
\end{equation*}
It is straightforward to verify that for every $s \geq s_1$,
\begin{equation} \label{eq:condition-on-rho-2}
2^{s}/r \geq 2 (s/2).
\end{equation}
This follows for $s_1$ because $2^{s_1} = \rho r \log(en/r)$ and
$$
\frac{2^{s_1}}{r} = (10\log_2 e) \cdot  \max\left\{\log\left(\frac{en}{r}\right),\log(er)\right\} \geq s_1,
$$
and for $s>s_1$ because $s \mapsto 2^s/s$ is increasing. Therefore,
$$
\sup_{v \in V_r} \sum_{s \geq s_1} 2^{s/2} \|\Delta_s v\|_2 \leq c_0 \frac{n}{r} \cdot \sum_{s \geq s_1} 2^{-2^s/(2r)},
$$
which is dominated by a geometric series with power \\
$2^{-2^{s_1}/2r} = 2^{-(\rho/2)\log(en/r)} \leq 1/4$. Therefore,
\begin{equation*}
\frac{n}{r}\sum_{s \geq s_1} 2^{-2^s/2r} \lesssim  \frac{n}{r} \cdot  2^{-2^{s_1}/(2r)}
\leq e^{-1} \left(\frac{r}{en}\right)^{\frac{\rho}{2\log_2 e}-1}.
\end{equation*}
Note that
\begin{equation} \label{eq:condition-on-rho}
\left(\frac{r}{en}\right)^{\frac{\rho}{2\log_2 e}-1} \leq \frac{1}{(en)^2}.
\end{equation}
Indeed, if $r \leq n/r$, i.e., if $r \leq \sqrt{n}$, then $\rho/\log_2 e \geq 10$ and
$$
\left(\frac{r}{en}\right)^{(\rho/2\log_2e)-1} \leq \left(\frac{1}{\sqrt{en}}\right)^{4} \leq \frac{1}{(en)^2};
$$
otherwise, $\sqrt{n} \leq r \leq n/2$ and $\rho/\log_2 e = 10\log(er)/\log(en/r)$ so that
\begin{equation*}
\log\left( \left(\frac{en}{r}\right)^{(\rho/2\log_2 e)-1}\right) = \log\left(\frac{en}{r}\right) \left[ \frac{5 \log(er)}{\log(en/r)}-1 \right]
\geq \log \left((en)^{2} \right),
\end{equation*}
so that \eqref{eq:condition-on-rho} holds also in this case.
Therefore,
\begin{equation*}
\sup_{v \in V_r} \sum_{s \geq s_1} (\sqrt{n}+\sqrt{r} 2^{s/2}) \|\Delta_s v\|_2 \leq \frac{c_1}{n^{3/2}}
\end{equation*}
for an absolute constant $c_1$.
\endproof

\section{The structure of a typical $\{\Gamma_v \xi : v \in V_r\}$}
\label{sec:main-technical}
The main component in the proof of Theorem \ref{thm:iid-subgaussian} is a structural result on a typical realization of the random set $\{\Gamma_v \xi : v \in V_r\}$.

\begin{thm} \label{thm:main-structure}
Recall the definitions of $\rho, s_0, s_1,$ and $\alpha_r$ from $\eqref{eq:rho}, \eqref{eq:s},$ and $\eqref{eq:alpha}$, respectively.  There exist constants $c_0,\hdots,c_3$ that depend only on $L$ and $\beta$ for which the following holds.
With probability at least $1-2^{-c_0 \min\{2^{s_0},2^{s_1}\}}$,
$$
\{\Gamma_v \xi : v \in V_r\} \subset T_1 + T_2,
$$
where
$$
|T_1| \leq 2^{2^{s_1}} \ \ {\rm and} \ \ T_2 \subset c_1n^{-3/2}B_2^n.
$$
Moreover, for each $t \in T_1$, 
\begin{enumerate}
\item $\| t \|_2 \geq \kappa_1 \sqrt{n},$ and 
\item $\| t_{I} \|_2 \leq \frac{\kappa_1 \sqrt{n}}{4}$ on the set $I=I(t) \subset [n]$ of largest-magnitude coordinates of $t$ of cardinality $|I| \geq c_2 \frac{n}{\alpha_r^2 \log(e \alpha_r)}.$ 
\end{enumerate}

\end{thm}


Theorem \ref{thm:main-structure} implies that a typical realization of $\{\Gamma_v \xi : v \in V_r\}$ is just a perturbation of the (random) set $T_1$, and that $T_1$ consists of vectors with a regular coordinate structure: 
{for each $t \in T_1$ consider the set $J = I(t)^c$ and observe that it follows from $\| t \|_2 \geq \kappa_1 \sqrt{n}$
and $\| t_{I} \|_2 \leq \frac{\kappa_1 \sqrt{n}}{4}$ that $\| t_J\|_2 \geq \frac{15 \kappa_1}{16} \sqrt{n}$ and
\begin{equation}\label{tJ_bound}
\|t_J\|_\infty \leq |t^*_{|I(t)|}| \leq \left( \frac{1}{|I(t)|} \sum_{j \in I(t)} |t_j^*|^2\right)^{1/2} \leq \frac{\kappa_1}{4\sqrt{c_2}} \alpha_r \sqrt{\log(e \alpha_r)}. 
\end{equation}
This implies that at least $\theta=c/(\alpha_r^2 \log(e \alpha_r))$ coordinates of $t_J$ are larger than a constant.}
When $s_0 \geq  s_1$ (the low-sparsity case), $\alpha_r=1$, and the claim is that with probability at least $1-2^{-c_0\rho r \log(en/r)}$, for every $t \in T_1$, there is a subset $I \subset [n]$ of cardinality $|I| \geq c_2 n$ of coordinates which are larger than a constant, and thus, each one of the vectors $t \in T_1$ has a regular coordinate structure in the sense of \eqref{eq:reg-vector}.

When $s_1 > s_0$, a similar type of claim holds, but with probability at least $1-2^{-c_0\kappa_4 n/r}$, and the regularity condition on the coordinates of $t \in T_1$ is slightly weaker: one no longer has a subset of cardinality that is proportional to $n$ consisting of coordinates that are larger than a constant, but rather a (marginally) smaller set $I$.

Thanks to this information on the structure of a typical $\{\Gamma_v \xi : v \in V_r\}$, one may establish the required lower bound on $\inf_{v \in V_r} \|P_\Omega \Gamma_v \xi\|_2$.

\begin{Corollary} \label{cor:from-structure-to-proj}
There exist constants $c_0$, $c_1$ and $c_2$ that depend only on $L$ and $\beta$ for which the following holds. Let
$$
\delta n \geq c_0 (\alpha_r^2 \log(e \alpha_r)) \cdot  \rho r\log(en/r),
$$
with $\rho$ defined in \eqref{eq:rho} and set $(\delta_i)_{i=1}$ to be independent selectors with mean $\delta$.
Then with probability at least $1-2^{-c_1 \min\{2^{s_0},2^{s_1}\}}$,
$$
\inf_{v \in V_r} \sum_{i=1}^n \delta_i \inr{\Gamma_v \xi,e_i}^2 \geq c_2 \delta n.
$$
\end{Corollary}

\proof
Let $\xi$ be a realization of the event from Theorem \ref{thm:main-structure} -- which holds with probability at least $1-2^{-c_0\min\{2^{s_0},2^{s_1}\}}$ relative to $\xi$ {--} and let $(\delta_i)_{i=1}^n$ be independent selectors with mean $\delta$ that are also independent of $\xi$.

Using the notation of Theorem \ref{thm:main-structure}, let $\Gamma_v \xi = t+y$ for $t \in T_1$ and $y \in T_2$; hence
\begin{align}
\left(\sum_{i=1}^n \delta_i \inr{\Gamma_v \xi,e_i}^2\right)^{1/2} &\geq \inf_{t \in T_1} \left(\sum_{i=1}^n \delta_i t_i^2\right)^{1/2} - \sup_{y \in T_2} \left(\sum_{i=1}^n \delta_i y_i^2\right)^{1/2} \nonumber \\
&\geq \inf_{t \in T_1} \left(\sum_{i=1}^n \delta_i t_i^2\right)^{1/2} - \frac{c_1}{n^{3/2}} \label{low_bound_split}
\end{align}
for an absolute constant $c_1$. It suffices to show that
$$
\inf_{t \in T_1} \sum_{i=1}^n \delta_i t_i^2 \geq c_2 \delta n, 
$$
the right hand being larger than $2 c_1/n^{3}$, where $c_2$ is a suitable constant.

\medskip

{
Fix $t \in T_1$, let $J=J(t)=I^c(t)$ be the {\emph{complement}} of the set $I(t)$ identified by Theorem \ref{thm:main-structure} and put $x=P_J t$. Then} {as argued in \eqref{tJ_bound} and the line above  
\begin{equation} \label{eq:cond-on-x}
\| x \|_2 \geq \frac{15c_1}{16} \sqrt{n} \quad \mbox{ and } \quad
\| x \|_{\infty} \leq 
c_3 \alpha_r \sqrt{\log(e \alpha_r)}.
\end{equation}}
By Bernstein's inequality,
$$
Pr \Bigl( \bigl|\sum_{i \in J} (\delta_i-\delta)x_i^2 \bigr| \geq w \Bigr) \leq 2\exp\Bigl(-c_4\min\Bigr\{\frac{w^2}{\delta \sum_{i \in J}x_i^4},\frac{w}{\max_{i \in J} x_i^2} \Bigr\}\Bigr).
$$
Observe that $\sum_{i \in J} x_i^4 \leq \|x\|_\infty^2 \sum_{i \in J} x_i^2$, and thus, for $w = (\delta/2) \sum_{i \in J} x_i^2$, the probability estimate becomes
$$
2\exp(-c_4\delta \|x\|_2^2/\|x\|_\infty^2) \leq 2\exp\Bigl(-c_5(L,\beta)  \frac{\delta n}{\alpha_r^2 \log(e\alpha_r)} \Bigr).
$$
By a union bound, with probability at least
$$
1-2|T_1|2\exp\Bigl(-c_5  \frac{\delta n}{\alpha_r^2 \log(e\alpha_r)} \Bigr),
$$
relative to $(\delta_i)_{i=1}^n$, this implies that, for every $t \in T_1$,
\begin{equation*}
\sum_{i=1}^n \delta_i t_i^2 \geq \sum_{i \in J(t) } \delta_i t_i^2 \geq \frac{\delta}{2} \sum_{i \in J(t)} x_i^2 \geq \frac{\kappa_1^2}{32} \delta n.
\end{equation*}
The claim follows by setting
$$
\delta n \geq c_5^{-1} (\alpha_r^2 \log(e\alpha_r)) \cdot 2^{s_1+1} = c_7(L,\beta) \cdot \alpha_r^2 \log (e\alpha_r) \cdot \rho r \log(en/r).
$$
\endproof

{{
\begin{Remark} The following alternative argument starting from \eqref{eq:cond-on-x} leads to a lower bound
in $\ell_q$ for any $1 \leq q < \infty$ -- at least in the small sparsity regime $s_0 \leq s_1$, i.e., $s \leq c \sqrt{n/\log(n)}$.
For fixed $t \in T_1$ choose $I = I(t)$ as the $\theta n$ largest absolute coefficients of $t$ where $\theta$ is defined in \eqref{def:theta} below (with suitable constants). We know from Theorem~ that $\|t\|_2 \geq \kappa_1 \sqrt{n}$ and $\|t\|_{[\theta n]} \leq \frac{\kappa}{4}\sqrt{n}$. This implies that
\begin{align*}
t^*_{\theta n} & \geq \left( \frac{1}{(1-\theta)n} \sum_{i = \theta n}^n (t_i^*)^2 \right)^{1/2}
= \left( \frac{1}{(1-\theta)n} \left( \|t\|_2^2 - \|t\|_{[\theta n]}^2\right)\right)^{1/2} \\
& \geq \left( \frac{1}{n}(\kappa_1^2 n - \kappa_1^2n/16)\right)^{1/2}
= \frac{\sqrt{15}\kappa_1}{4} = c_6.
\end{align*}
For $\Omega = \{i :\delta_i = 1\}$ let $K=K(t) = I(t) \cap \Omega$. Then, by Chernov's inequality 
the cardinality of $K$ satisfies $|K | = \sum_{j \in I(t)} \delta_i \geq \frac{1}{2} \delta | I | = \frac{1}{2} \delta \theta n$ 
probability at least $1- e^{-\delta \theta n/8}$. On this event,
\[
\|P_\Omega t \|_q \geq \left( \sum_{j \in K} |t_j|^q\right)^{1/q} \geq c_6 (\theta \delta n/2)^{1/q}
\]
By the union bound, this holds for all $t \in T_1$ with probability at least
\begin{align*}
1 - 2 |T_1| \exp(-\delta \theta n) & \geq 1 - 2\cdot 2^{2^{s_1}} \exp(-\delta \theta n/8) \\
& \geq 1 - 2 \exp( -\delta \theta n/8 + \ln(2) \rho r \log(en/r)).
\end{align*}
Therefore, in the small sparsity regime $s_0 \leq s_1$, i.e., $\theta=c_3$, a combination with a similar estimate
as in \eqref{low_bound_split} gives
\[
\inf_{v \in V_r} \sum_{i=1}^n \delta_i \langle \Gamma_v \xi,e_i\rangle^q \geq c_8^q  \delta n  
\]
with probability at least $1-2 \exp(-c_9 \delta n)$ provided that 
\[
\delta n \geq c_{10} r \log(en/r).
\]
\end{Remark}}
}
\endproof

The proof of Theorem \ref{thm:main-structure} is based on the following idea: $(V_{r,s})_{s \geq s_1}$ will be selected as an maximal $\eps_s$-separated subset 
of $V_r$ and $T_1=\{\Gamma_v \xi : v \in V_{r,s_1}\}$. We will show that for every $v \in V_r$, $v-\pi_{s_1} v = \sum_{s \geq s_1} \Delta_s v$ is 
small enough to ensure that with high probability,
$$
\sup_{v \in V_r} \|\Gamma_{v-\pi_{s_1}v}\xi\|_2 \leq \sum_{s \geq s_1} \|\Delta_s v\|_2 \leq \frac{c^\prime}{n^{3/2}}
$$
for $c^\prime$ that depends on $L$ and $\beta$. Then, we will turn to the more difficult part of the argument -- that with high probability, for every $v \in V_{r,s_1}$,
\begin{equation} \label{bound:regular}
\|\Gamma_{v} \xi\|_2 \geq c\sqrt{n} \ \ {\rm and} \ \ \|\Gamma_v \xi\|_{[\theta n]} \leq (c/4)\sqrt{n}
\end{equation}
for well-chosen $c$ and 
\begin{equation}\label{def:theta}
\theta = \left\{ \begin{array}{cc} c_1 & \mbox{ for } r \leq c_2 \sqrt{ \frac{ \kappa_4 n}{\log( c_2n/\kappa_4)}} \\
\frac{c_3}{\alpha_r^2 \log(e \alpha_r)} & \mbox{ for } c_2 \sqrt{ \frac{ \kappa_4 n}{\log( c_2n/\kappa_4)}} \leq r \leq \frac{c_4 n}{\log^4(n)}
\end{array} \right.
\end{equation}
where all constants $c,c_1,c_2,c_3,c_4$ only depend on $L$ and $\beta$ and $\kappa_4$ is defined in \eqref{def:kappa4},
see \eqref{eq:k-norm-centres-low-sparse} and the following remarks as well as Lemma~\ref{lem:high-sparsity}. 
Moreover, here and in the following, we assume for simplicity that $\theta n$ is an integer. (The general case may need slightly different constants.)

We begin the proof with its simpler part: a high probability estimate on $\sup_{v \in V_r} \|\Gamma_{v-\pi_{s_1}v}\xi\|_2$.

\begin{Lemma} \label{lemma:osc-of-Gamma-V}
There exist constants $c$ and $c_1$ that depend only on $L$ and $\beta$ for which the following holds.
If $(V_{r,s})_{s \geq s_1}$ is an admissible sequence of $V_r$, then with probability at least $1-2^{-2^{s_1}}$, for every $v \in V_r$ and $s \geq s_1$,
\begin{equation} \label{eq:oscillations}
\|\Gamma_{\Delta_s v} \xi\|_{2} \leq c\left(\sqrt{n} + 2^{s/2}\sqrt{r}\right)\|\Delta_s v\|_2.
\end{equation}
In particular, if 
{$V_{r,s_1}$} is a maximal separated subset of $V_r$, then with probability at least $1-2^{-2^{s_1}}$,
$$
\sup_{v \in V_r} \|\Gamma_{v-\pi_{s_1}v} \xi\|_2 \leq \frac{c_1}{n^{3/2}}.
$$
\end{Lemma}

\proof
The first part of the proof is a straightforward outcome of \eqref{eq:basic-estimate-Gamma-v-2}. Indeed, if we set $k=n$ and $p=2^{s+3}$, then by Corollary~\ref{cor:basic-estimates-Gamma-v}, with probability at least $1-\exp(-{2^{s+3}})$,
$$
\|\Gamma_{\Delta_s v} \xi\|_{2} \leq 4\kappa_3 (\sqrt{n}\|\Delta_s v\|_2 + 2^{s/2} \cdot \sqrt{n} \|W\Delta_s v\|_\infty).
$$
Since $v$ is $r$-sparse, $\Delta_s v = \pi_{s+1}v-\pi_s v$ is $2r$-sparse, and since $W$ is a Hadamard type matrix with constant $\beta$,
$$
\sqrt{n}\|W\Delta_s v\|_\infty = \sqrt{n} \max_{i \in [n]} |\inr{W_i,\Delta_s v}| \leq \beta \|\Delta_s v\|_1 \leq \beta \|\Delta_s v\|_0^{1/2} \cdot \|\Delta_s v\|_2.
$$
Therefore,
\begin{equation} \label{eq:chaining-large-s-1}
\|\Gamma_{\Delta_s v} \xi\|_{2} \leq {4}\kappa_3 (\sqrt{n} + \beta \sqrt{{2}r}\, 2^{s/2})\|\Delta_s v\|_2.
\end{equation}
There are at most $2^{2^s}\cdot 2^{2^{s+1}} \leq 2^{2^{s+2}}$ vectors of the form $\Delta_sv$, and thus, by a union bound,
with probability at least $1-2^{-2^{s+2}}$, \eqref{eq:chaining-large-s-1} holds for every $v \in V_r$ for that choice of $s$. Summing the probabilities for $s \geq s_1$, one has that with probability at least $1-2^{-2^{s_1+1}}$, for every $v \in V_r$,
\begin{align*}
\|\Gamma_{v-\pi_{s_1}v}\xi\|_2 &=  \|\sum_{s \geq s_1} \Gamma_{\Delta_s v}\xi\|_2 \leq \sum_{s \geq s_1} \|\Gamma_{\Delta_s v}\xi\|_2  \nonumber \\
&\leq {4\sqrt{2}} \kappa_3 \beta \sum_{s \geq s_1} \left(\sqrt{n} + \sqrt{r} 2^{s/2}\right)\|\Delta_s v\|_2 \leq \frac{c(L,\beta)}{n^{3/2}},
\end{align*}
where the last inequality is just Lemma \ref{lemma:covering-sphere}.
\endproof

Lemma \ref{lemma:osc-of-Gamma-V} shows that if we set $T_2=\{\Gamma_{v-\pi_{s_1}v} \xi : v \in V_r\}$,
then with probability at least $1-2^{-2^{s_1}}$,
$\sup_{t \in T_2} \|t\|_2 = \sup_{v \in V_r} \|\Gamma_{v-\pi_{s_1}v} \xi\|_2 \leq c(L,\beta)/n^{3/2}$,
as required in Theorem \ref{thm:main-structure}. Therefore, all that is left is to study the structure of $T_1 =\{\Gamma_{\pi_{s_1}v}\xi : v \in V_r\}$, and to show that with high probability, it consists of vectors with a regular coordinate structure. To that end we shall split the argument into two cases: the low-sparsity case, when $s_0 \geq s_1$ and the high-sparsity one, in which the reverse inequality holds. The analysis {is based on Corollary \ref{cor:basic-estimates-Gamma-v} in both cases}.

\subsubsection*{The low-sparsity case}

Assume that $2^{s_0} \geq 2^{s_1}${{,i.e.,} $\frac{\kappa_4 n}{r} \geq \rho r \log(en/r)$ and in particular, by the choice of $\kappa_4$ {in \eqref{def:kappa4}},
\begin{equation} \label{eq:condition-on-lambda}
\frac{\kappa_2}{2} \cdot \frac{n}{r} \geq  \rho r \log\left(\frac{en}{r}\right).
\end{equation}
Fix $v \in V_{r,s_1}$. It follows from \eqref{eq:basic-estimate-Gamma-v-1} that with probability at least $1-2^{-\kappa_2n/r}$,
\begin{equation} \label{eq:2-norm-centres-low-sparse}
\|\Gamma_v \xi\|_2 \geq \kappa_1 \sqrt{n}.
\end{equation}
Let $0<\theta<1$ to be named later. Observe that by \eqref{eq:basic-estimate-Gamma-v-2} for $k=\theta n$ and $p=2\rho r \log(en/r)=2^{s_1+1}$, with probability at least $1-{e}^{-2^{s_1+1}}$,
\begin{align} \label{eq:k-norm-centres-low-sparse}
\|\Gamma_v \xi\|_{[k]} \leq & 2\kappa_3(\|v\|_2 \sqrt{k\log(en/k)} + \sqrt{\rho r \log(en/r)} \sqrt{n} \|Wv\|_\infty) \nonumber
\\
\leq & 2\kappa_3 (\sqrt{n}\sqrt{\theta \log(e/\theta)} + \beta \sqrt{\rho r \log(en/r)} \cdot \sqrt{r}),
\end{align}
because $\|v\|_2=1$ and $\sqrt{n}\|Wv\|_\infty \leq \beta \sqrt{r}$.

Recall that $|V_{r,s_1}| \leq 2^{2^{s_1}}$ and thus, by \eqref{eq:condition-on-lambda} and the union bound,  with probability at least $1-2^{-2^{s_1}}$, \eqref{eq:2-norm-centres-low-sparse} and \eqref{eq:k-norm-centres-low-sparse} hold for every $v \in V_{r,s_1}$. All that remains is {to} ensure that
$$
2\kappa_3 \beta \sqrt{\theta \log(e/\theta)} \sqrt{n} \leq \frac{\kappa_1}{{8}} \sqrt{n} \ \ \ {\rm and} \ \ \ 2\kappa_3 \beta \sqrt{r} \cdot \sqrt{\rho r \log(en/r)} \leq \frac{\kappa_1}{{8}} \sqrt{n}.
$$
The first condition holds for the right choice of the constant $\theta=\theta(L,\beta)$. For the second, note that by the definition
of $\kappa_4$ in \eqref{def:kappa4}
$$
\rho r \log(en/r) =2^{s_1} \leq 2^{s_0} \leq \frac{\kappa_1^2}{{256} \kappa_3^2 L^2 \beta^2} \cdot \frac{n}{r}.
$$
Therefore,
$$
2\kappa_3 \beta \sqrt{r} \cdot \sqrt{\rho r \log(en/r)} \leq 2\kappa_3 \beta \sqrt{r} \cdot \frac{\kappa_1}{{16} \kappa_3 L \beta} \cdot \sqrt{\frac{n}{r}} \leq \frac{\kappa_1}{{8}} \sqrt{n},
$$
because $L \geq 1$.

This concludes the proof of Theorem \ref{thm:main-structure} in the low sparsity case.
\endproof

\subsubsection*{The high-sparsity case}
Now consider the case $2^{s_0} \leq 2^{s_1}${, that is,} $\kappa_4 n/r \leq \rho r \log(en/r)$. {Then} there is a substantial gap between the individual probability estimate \eqref{eq:basic-estimate-Gamma-v-1}
and the cardinality of $V_{r,s_1}$, so that a simple union bound does not give a non-trivial probability estimate. The difficulty one faces here is bridging this gap, and the key to the proof in the high-sparsity case is the following result.
\begin{thm} \label{thm:bridging}
There exist constants $c_1$ and $c_2$ that depend only on $L$ and $\beta$ for which the following holds. If $r \leq c_1 n/\log^4n$ then with probability at least $1-2\exp(-c_2n/r)$,
\[
eq{\inf_{v \in V_{r,s_1}} \|\Gamma_v \xi\|_2 \geq \frac{\kappa_1}{2}\sqrt{n}}.
\]
\end{thm}

For the proof of Theorem \ref{thm:bridging}, one has to `transfer' the lower bound on $\inf_{v \in V_{r,s_0}} \|\Gamma_v \xi\|_2$, which may be obtained directly from \eqref{eq:basic-estimate-Gamma-v-1} and the union bound, to the much larger set $V_{r,s_1}$. Thus, it suffices to show that with high probability,
\begin{equation} \label{eq:trans-cond}
\sup_{v \in V_{r,s_1}} \left| \|\Gamma_v \xi\|_2^2 - \|\Gamma_{\pi_{{s_0}} v}\|_2^2 \right| \leq \frac{\kappa_1^2}{4} n,
\end{equation}
for an approximating $\pi_{{s_0}} v \in V_{r,s_0}$.

To address \eqref{eq:trans-cond}, we proceed along the lines of \cite{krmera14} and
consider the following, more general situation: let ${\cal A}$ be a class of matrices, $|{\cal A}| \leq 2^{2^{s_1}}$ and set $({\cal A}_s)_{s \geq s_0}$ to be an admissible sequence of ${\cal A}$; that is, ${\cal A}_{s}$ is of cardinality at most $2^{2^{s}}$. Let $\pi_s A $ to be the nearest point to $A$ in ${\cal A}_s$  with respect to the $\| \ \|_{2 \to 2}$ norm, set $\Delta_s A = \pi_{s+1}A-\pi_s A$, and put
$$
\gamma_{s_0,s_1}({\cal A})=\sup_{A \in {\cal A}} \sum_{s=s_0}^{s_1-1} 2^{s/2}\|\Delta_s A\|_{2 \to 2}.
$$

\begin{Lemma} \label{lemma:from-KMR}
{Assume $\mathcal{A} = \mathcal{A}_{s_1}$ to be finite.}
Let $\xi$ be an isotropic, $L$-subgaussian random vector and set $\xi^\prime$ to be an independent copy of $\xi$. Let $N_{\cal A}(\xi) = \sup_{A \in {\cal A}} \|A\xi\|_2$ and put
\[
{{\cal Z} = \sup_{A \in {\cal A}} \left| \inr{A\xi,A\xi^\prime} - \inr{(\pi_{s_0} A)\xi,(\pi_{s_0} A) \xi^\prime} \right|}.
\]
Then, for every $p \geq 1$,
$$
\|{\cal Z}\|_{L_p} \leq cL \sup_{A \in {\cal A}}\gamma_{s_0,s_1}({\cal A}) \|N_{\cal A}(\xi)\|_{L_p}.
$$
for an absolute constant $c$.
\end{Lemma}
Since the proof of Lemma \ref{lemma:from-KMR} is contained in the proof of Lemma 3.3 in \cite{krmera14}, it will not be presented here.

We will be interested in the specific class of matrices ${\cal A}=\{\Gamma_v : v \in V_{r,s_1}\}$:
\begin{thm} \label{thm:transfer-norm control}
There exist constants $c_1$ and $c_2$ that depend only on $L$ and $\beta$ for which the following holds. Let ${\cal A}=\{\Gamma_v : v \in V_{r,s_1}\}$ and set $({\cal A}_s)_{s \geq s_0}$ to be an admissible sequence of ${\cal A}$. Then, with probability at least $1-2^{-c_1n/r}$,
\begin{equation}
\sup_{A \in {\cal A}}\left|\|A \xi\|_2^2 - \|\pi_{s_0} A \xi\|_2^2\right| \leq c_2\gamma_{s_0,s_1}({\cal A}) \left(\gamma_{s_0,s_1}({\cal A})+\sqrt{n}\right).
\end{equation}
\end{thm}

Before proving Theorem \ref{thm:transfer-norm control}, let us recall a standard fact that can be established using tail integration.
\begin{Lemma} \label{lemma:prob-obs}
Let $Z$ be a nonnegative random variable and assume that
$$
Pr(Z \geq A_1 + u A_2) \leq 2\exp(-u^2/2) \ \ \ \ {\rm for \ every \ } u \geq A_3.
$$
Then, for every $p \geq 1$, $\|Z\|_{L_p} \leq c(A_1 + A_2 \cdot \max\{A_3,\sqrt{p}\})$, and $c$ is an absolute constant.
\end{Lemma}

\noindent{\bf Proof of Theorem \ref{thm:transfer-norm control}.}
Denote by $A^j$, $j\in [n]$, the columns of the matrix $A$ and observe that
\begin{align*}
\|A \xi\|_2^2 - \|\pi_{s_0} A \xi\|_2^2 &= \inr{A\xi,A\xi}^2 - \inr{\pi_{s_0}A\xi,\pi_{s_0}A\xi}^2 \nonumber \\
&= \sum_{j,k} \xi_j \xi_k \left(\inr{A^j,A^k} - \inr{(\pi_{s_0}A)^j,(\pi_{s_0}A)^k}\right).
\end{align*}
Since each $A$ is of the form $\Gamma_v$ for some $v \in S^{n-1}$,  we have $\inr{A^j,A^j}=\|v\|_2^2=1$, and the same holds for $\pi_{s_0}A$. Therefore, $\inr{A^j,A^j} - \inr{(\pi_{s_0}A)^j,(\pi_{s_0}A)^j}=0$ and all that remains is to control the ``off-diagonal" terms,
$$
\sum_{j \not=k} \xi_j \xi_k \left(\inr{A^j,A^k} - \inr{(\pi_{s_0}A)^j,(\pi_{s_0}A)^k}\right) =: F_A.
$$
{Denoting by $\xi'$ an independent copy of $\xi$}, applying a standard decoupling argument (see, for example, \cite[Theorem 2.4]{krmera14} or \cite{gide99})
and Lemma~\ref{lemma:from-KMR}
\begin{align*}
(\E \sup_{A \in {\cal A}} |F_A|^p)^{1/p} & =   \|\sup_{A \in {\cal A}} (*)_A\|_{L_p} \nonumber \\
&\leq {4} \left\| \sup_{A \in {\cal A}} \left|\sum_{j,k} \xi_j \xi^\prime_k \left(\inr{A^j,A^k} - \inr{(\pi_{s_0}A)^j,(\pi_{s_0}A)^k} \right) \right| \right\|_{L_p}
\\
& =  {4} \left\| \sup_{A \in {\cal A}} \left| \inr{A\xi,A\xi^\prime} - \inr{\pi_{s_0}A\xi,\pi_{s_0} A\xi^\prime} \right| \right\|_{L_p}
\\
& \leq {c_1} L \gamma_{s_0,s_1}({\cal A}) \cdot \|N_{\cal A}(\xi)\|_{L_p}.
\end{align*}
Moreover,
\begin{align*}
\|(N_{\cal A}(\xi))^2\|_{L_p} \leq & \left\| \sup_{A \in {\cal A}} \left| \|A \xi\|_2^2 - \|\pi_{s_0} A \xi\|_2^2 \right| \right\|_{L_p} + \left\| \sup_{A \in {\cal A}} \|\pi_{s_0} A \xi\|_2^2 \right\|_{L_p}
\\
\leq & {c_1} L \gamma_{s_0,s_1}({\cal A}) \cdot \|N_{\cal A}(\xi)\|_{L_p} + \left\| \sup_{A \in {\cal A}} \|\pi_{s_0} A \xi\|_2^2 \right\|_{L_p}.
\end{align*}
By \eqref{eq:basic-estimate-Gamma-v-2} for $k=n$ and $p=u^2 \geq 2^{s_0+3}$, it follows that with probability at least $1-2\exp(-u^2/2)$,
\begin{align*}
 \sup_{A \in {\cal A}} \|\pi_{s_0} A \xi\|_2 & = \sup_{v \in V_{r,s_0}} \|\Gamma_v \xi\|_2 \leq \kappa_3(\sqrt{n}+u \sqrt{n}\|Wv\|_\infty) \\
&  \leq {c_2}(L,\beta) (\sqrt{n} + u \sqrt{r}),
\end{align*}
where we used that $\|Wv\|_\infty \leq \beta \sqrt{r/n}$.
Thus, the random variable $Z=\sup_{A \in {\cal A}} \|\pi_{s_0} A \xi\|_2$ satisfies the conditions of Lemma \ref{lemma:prob-obs} for $A_1 = c_3 \sqrt{n}$, $A_2=c_3\sqrt{r}$ and $A_3=4 \cdot 2^{s_0/2}$, implying that for every $p \geq 1$,
$$
\left\| \sup_{A \in {\cal A}} \|\pi_{s_0} A \xi\|_2^2 \right\|_{L_p} \leq (\E Z^{2p})^{1/p} \leq c_4(L,\beta) (\sqrt{n} + \sqrt{r} \max\{2^{s_0/2},\sqrt{p}\})^2.
$$
Setting $p=2^{s_0} = \kappa_4 n/r$ we obtain
$$
\left\| \sup_{A \in {\cal A}} \|\pi_{s_0} A \xi\|_2^2 \right\|_{L_p} \leq c_5(L,\beta)n.
$$
{Therefore},
$$
\|(N_{\cal A}(\xi))^2\|_{L_p} \leq c_6(L,\beta) \left(\gamma_{s_0,s_1}({\cal A}) \|N_{\cal A}(\xi)\|_{L_p} + n\right),
$$
implying that
$$
\|N_{\cal A}(\xi)\|_{L_p} \leq c_7(L,\beta)\max\{\gamma_{s_0,s_1}({\cal A}),  \sqrt{n} \}
$$
and
\begin{equation*}
\Bigl(\E \sup_{A \in {\cal A}}\left|\|A \xi\|_2^2 - \|\pi_{s_0} A \xi\|_2^2\right|^p\Bigr)^{1/p} \leq c_8(L,\beta)\gamma_{s_0,s_1}({\cal A}) \left(\gamma_{s_0,s_1}({\cal A})+\sqrt{n}\right).
\end{equation*}
The claim now follows from Lemma~\ref{lem:moments} with $\alpha=2$ and the definition of $p=\kappa_4 n/r$.
\endproof

The next step is to estimate $\gamma_{s_0,s_1}({\cal A})$ for our choice ${\cal A}=\{\Gamma_v : v \in V_{r,s_1}\}$. We will construct the admissible sequence ${\cal A}_s=\{\Gamma_v : v \in V_{r,s}\}$ for $s_0 \leq s <s_1$ based on the fact that
$$
\|\Delta_s A\|_{2 \to 2} = \|\Gamma_{\pi_{s+1}v}-\Gamma_{\pi_{s+1}v}\|_{2 \to 2} = \sqrt{n} \|W \Delta_s v\|_\infty,
$$
for $\Delta_s v=\pi_{s+1}v -\pi_s v$. Hence,
$$
\gamma_{s_0,s_1}({\cal A})= \sup_{v \in V_{r,s_1}} \sum_{s=s_0}^{s_1-1} 2^{s/2} \cdot \sqrt{n} \|W\Delta_s v\|_\infty,
$$
and the admissible sequence will be constructed as maximal separated subsets of  $V_{r,s_1}$ with respect to the norm $\sqrt{n} \|W(\cdot)\|_\infty$.

We will require a well-known fact, due to Carl \cite{ca85}.
\begin{Lemma} \label{lemma:carl}
There is an absolute constant $c$ for which the following holds. Let $w_1,\hdots,w_n \in \R^n$ that satisfy $\|w_i\|_\infty \leq K$, put $\|z\|=\max_{i \in [n]} |\inr{z,w_i}|$ and set $B$ to be the unit ball with respect to that norm. Then for every $t >0$,
$$
\log N(\sqrt{r} B_1^n, tB) \leq C\frac{K^2r}{t^2} \log^2(nt^2/r).
$$
\end{Lemma}
In the case we are interested in, $w_i = \sqrt{n}W_i$ and $\|w_i\|_{\infty} \leq \beta$. Moreover, $V_{r,s_1} \subset V_r \subset \sqrt{r}B_1^n$; therefore,
\begin{equation} \label{eq:infty-cover-est}
\log N(V_{r,s_1}, tB) \leq \log N(\sqrt{r}B_1^n, tB) \leq
C\frac{\beta^2 r}{t^2} \log^2 (nt^2/r).
\end{equation}
\begin{Corollary} \label{cor:admissible-infty}
There is an admissible sequence of $V_{r,s_1}$, for which, for every $v \in V_{r,s_1}$,
$$
\sqrt{n} \|W \Delta_s v\|_\infty \leq c \beta  2^{-s/2}\sqrt{r} \log(en/2^s) \quad {\mbox{ for } s \leq s_1,}
$$
for an absolute constant $c$. Therefore,
\begin{equation} \label{eq:gamma-wrt-infty}
\sup_{v \in V_{r,s_1}} \sum_{s=s_0}^{s_1-1} 2^{s/2} \cdot \sqrt{n} \|W\Delta_s v\|_\infty \leq c_1\beta \sqrt{r} \alpha_r \log(er/\kappa_4).
\end{equation}
\end{Corollary}
Corollary \ref{cor:admissible-infty} follows from \eqref{eq:infty-cover-est}, a straightforward computation and the definition of $s_0$ and $s_1$. We omit the details.

\noindent{\bf Proof of Theorem \ref{thm:bridging}.}
Combining the individual small ball estimate in \eqref{eq:basic-estimate-Gamma-v-1}, Theorem \ref{thm:transfer-norm control} and Corollary \ref{cor:admissible-infty}, one has that with probability at least $1-2^{-c_1(L,\beta)n/r}$, for every $v \in V_{r,s_1}$,
$$
\|\Gamma_{\pi_{s_0}v}\xi\|_2 \geq \kappa_1 \sqrt{n},
$$
and
$$
\left|\|\Gamma_v \xi\|_2^2 - \|\Gamma_{\pi_{s_0}v} \xi\|_2^2\right| \leq c_2(L,\beta) \sqrt{n} \cdot \sqrt{r} \alpha_r \log(c_2r).  
$$
It is straightforward verify that the latter term is bounded by $(\kappa_1^2/4)n$ provided that $r \leq c_3(L,\beta)n/\log^4n$.
\endproof

The more difficult step consists in exposing the regular coordinate structure of vectors in a typical $\{\Gamma_v \xi : v \in V_{r,s_1}\}$, and we will do that by finding a suitable upper bound on
\begin{equation} \label{eq:k-monotone-V-r-s-1}
\sup_{v \in V_{r,s_1}} \|\Gamma_v \xi\|_{[k]} = \sup_{v \in V_{r,s_1}} \Bigl(\sum_{i \leq k} (\inr{\Gamma_v \xi,e_i}^*)^2\Bigr)^{1/2}.
\end{equation}
Specifically, the next result 
{bounds} the largest possible $k$ for which \eqref{eq:k-monotone-V-r-s-1} is smaller than $(\kappa_1/4)\sqrt{n}$. 

\begin{Lemma}\label{lem:high-sparsity} 
{Assume that $2^{s_0} \leq 2^{s_1}$, $r \leq c_1 n/\log^4(n)$ and}
\begin{equation}\label{cond:k}
k \leq c_2 \frac{n}{\alpha_r^2 \log(e \alpha_r)}.
\end{equation}
Then with probability at least $1 - 2^{-c_3 2^{s_0}}$ 
\begin{equation}\label{bound:Vrs1}
\sup_{v \in V_{r,s_1}} \|\Gamma_v \xi\|_{[k]} \leq \frac{\kappa_1}{4} \sqrt{n}.
\end{equation}
The constants $c_1,c_2,c_3 {>0}$ only depend on $L$ and $\beta$.
\end{Lemma}
\begin{Remark} One should note that an upper bound on the supremum in \eqref{bound:Vrs1} cannot be obtained via an individual estimate and the union bound. 
The subgaussian property of $\xi$ is enough to ensure that for a fixed $v$,
\begin{equation*}
(\E\|\Gamma_v \xi\|^p_{[k]})^{1/p} \leq cL \left(\|v\|_2\sqrt{k\log(en/k)} + \sqrt{p}\sqrt{n}\|Wv\|_\infty\right).
\end{equation*}
However, because one only knows that for $v \in V_r$, $\sqrt{n}\|Wv\|_\infty \leq \beta \sqrt{r}\|v\|_2$, individual tail estimates suffice 
for a uniform bound in $V_{r,s_0}$, but not in the much larger set $V_{r,s_1}$.
\end{Remark}
\proof 
Observe that
$$
\sup_{v \in V_{r,s_1}} \|\Gamma_v \xi\|_{[k]} = \sup_{x \in V_k} \sup_{v \in V_{r,s_1}} \inr{ \xi,\Gamma_v^* x}.
$$
We will study the supremum of the linear process $w \mapsto \inr{\xi,w}$ indexed by the set
$$
\{\Gamma_v^* x : v \in V_{r,s_1}, \ x \in V_k\}.
$$
Let $(V_{k,s})_{s \geq s_0}$ be an admissible sequence of $V_k$ which will be specified later on.

For $x \in V_k$, we consider $\pi_{s_1}x \in V_{k,s_1}$, whose cardinality is $2^{2^{s_1}}$, and $\pi_{s_0}x \in V_{k,s_0}$, whose cardinality is $2^{2^{s_0}}$ and write, for $v \in V_{r,s_1}$,
\begin{equation}\label{H:decomp}
\Gamma^*_v x = \underbrace{\Gamma^*_v (x-\pi_{s_1}x)}_{=:H_1} + \underbrace{(\Gamma^*_{v} \pi_{s_1}x -\Gamma^*_{\pi_{s_0}v} \pi_{s_0}x)}_{=: H_2} + 
\underbrace{\Gamma^*_{\pi_{s_0}v} \pi_{s_0}x}_{=: H_3}.
\end{equation}
While $H_3$ corresponds to the ``starting points" of every chain, the difference between $H_1$ and $H_2$ lies in the ``balance" between the contribution of $V_k$ and $V_{r,s_1}$ to each one of them. 
For $H_1$, there are $\sim 2^{2^{s_1}}$ points $v \in V_{r,s_1}$, but for an admissible sequence for $V_k$ one has
$$
\Gamma^*_v (x-\pi_{s_1}x) = \sum_{s \geq s_1} \Gamma^*_v (\pi_{s+1}x-\pi_s x) = \sum_{s \geq s_1} (\Gamma^*_v \Delta_s x),
$$
and for $s \geq s_1$, $|\{\Delta_s x : x \in V_k\}| \geq 2^{2^{s_1}} = |V_{r,s_1}|$. Hence, it is possible to treat {$H_1$} for each $v \in V_{r,s_1}$ separately {and apply a simple union bound}. 
In contrast, {the situation for {$H_2$} 
is ``more balanced", and requires a different argument.}

To deal with $H_1$ in \eqref{H:decomp}, {fix} $v \in V_{r,s_1}$ and {apply Lemma \ref{lemma:Klartag} for a standard Gaussian vector $G$ in $\R^n$ together with
\eqref{bound:Gammav:ei}} to obtain}
{
\begin{align*}
\E \sup_{t \in \Gamma_v^* V_k} \inr{G,t} = & \E\|\Gamma_v G\|_{[k]} = \E\Bigl(\sum_{i = 1}^k (\inr{G,\Gamma_v^*e_i}^*)^2 \Bigr)^{1/2}
\\
\leq & c \sqrt{k \log(en/k)} \max_{i \in [n]} \|\Gamma_v^* e_i\|_2 {\leq} c {\beta} \sqrt{k \log(en/k)},
\end{align*}}%
By the Majorizing Measures Theorem \cite{ta14-1}, there exists an admissible sequence of $V_k$ for which
$$
\sup_{x \in V_k} \sum_{s=0}^\infty 2^{s/2}\|\Gamma_v^* \Delta_s x\|_2 \sim \E \sup_{t \in \Gamma_v^* V_k} \inr{G,t} \leq c {\beta} \sqrt{k \log(en/k)}.
$$
Let us consider a part of that admissible sequence, namely, $(V_{k,s})_{s \geq s_1}$. As $\xi$ is an isotropic, $L$-subgaussian random vector, it follows that for every $s \geq s_1$ and every $x \in V_k$, with probability at least $1-2^{-2^{s+3}}$,
$$
|\inr{\xi,\Gamma_v^* \Delta_s x}| \leq cL2^{s/2} \|\Gamma_v^* \Delta_s x\|_2.
$$
Summing for $s \geq s_1$, one has that with probability at least $1-2^{-2^{s_1+2}}$, for every $x \in V_k$,
\begin{align} \label{eq:admis-for-V-k}
|\inr{\xi,\Gamma_v^*(x-\pi_{s_1}x)}| & \leq \sum_{s \geq s_1} |\inr{\xi,\Gamma_v^* \Delta_s x}| \leq cL\sum_{s \geq s_1} 2^{s/2} \|\Gamma_v^* \Delta_s x\|_2\nonumber\\
& \leq c {\beta} L\sqrt{k \log(en/k)} {\leq \frac{\kappa_1}{16} \sqrt{n}}
\end{align}
{{for a suitable choice of $c_2$ in \eqref{cond:k}.}}
Repeating this argument for every $v \in V_{r,s_1}$ and applying the union bound, one has that with probability at least $1-2^{-2^{s_1+1}}$, \eqref{eq:admis-for-V-k} holds for every $v \in V_{r,s_1}$.

Next, let us turn to $H_2$ in \eqref{H:decomp}. We will construct approximating subsets in the following way: let $(V_{r,s})_{s=s_0}^{s_1}$ be the admissible sequence of $V_{r,s_1}$ used earlier, consisting of maximal separated subsets with respect to the norm $\sqrt{n}\|W(\cdot)\|_{\infty}$, and put $\nu_s$ to be the mesh width of the net $V_{r,s}$.
For every $s_0 \leq s \leq s_1$ and $z \in V_{r,s}$, let $\| \ \|_z$ be the ellipsoid norm
$$
\|x\|_z^2 = \|D_{Wz}U^*x\|_2^2= n \sum_{\ell=1}^n \inr{W_\ell,z}^2 \inr{U^\ell, x}^2,
$$
and set $T_s(z)$ to be a maximal separated subset in $V_k$ with respect to $\| \ \|_z$, of cardinality $2^{2^{s}}$. Denote its mesh width by $\eps_s(z)$.  
Thus, for $s_0 \leq s < s_1$, $v \in V_{r,s+1}$ and $x \in V_k$,
\begin{equation} \label{eq:chaining-large-s}
\Gamma_{v} x = \Gamma^*_{v-v^\prime} x + \Gamma^*_{v^\prime} (x-x^\prime) + \Gamma^*_{v^\prime} x^\prime,
\end{equation}
where $v^\prime \in V_{s,r}$ satisfies $\sqrt{n}\|W(v-v^\prime)\|_\infty \leq \nu_s$, and $x^\prime \in V_k$ belongs to $T_s(v^\prime)$ -- the net of $V_k$ with respect to the norm $\| \ \|_{v^\prime}$, 
and fulfills
\begin{equation} \label{eq:eps-s}
\|\Gamma^*_{v^\prime}(x-x^\prime)\|_2 = \|x-x^\prime\|_{v^\prime} \leq \eps_s(v^\prime).
\end{equation}
Moreover, the cardinality of $\bigcup_{v^\prime \in V_{r,s}} T_s(v^\prime)$ is at most $2^{2^{s}}\cdot 2^{2^{s}} \leq 2^{2^{s+1}}$.

The required estimate on $H_2$ follows {from \eqref{eq:gamma-wrt-infty}}
and {once we} control $\eps_s(z)$ with respect to each one of the $\| \ \|_z$ norms. Indeed, {using the above notation}
\[
{|\inr{\xi,\Gamma^*_{v} x} - \inr{\xi,\Gamma^*_{\pi_s v} \tilde{\pi}_s x}}| \leq |\inr{\xi,\Gamma^*_{v-\pi_s v} x}| + |\inr{\xi,\Gamma^*_{\pi_s v} (x-\tilde{\pi}_s x)}|, 
\]
{where, for $v \in V_{r,s+1}$, $\pi_s v \in V_{r,s}$ is closest to $v$ with respect to $\sqrt{n} \|W(\cdot)\|_\infty$, 
$x \in T_{s+1}(v)$ and $\tilde{\pi}_s x \in T_s(\pi_s v)$ is closest to $x$ with respect to $\| \ \|_{\pi_s v}$}. Since $\xi$ is isotropic and $L$-subgaussian, we have with probability at least $1-2^{2^{s+2}}$, 
for every $v \in V_{r,s+1}$ and $x \in T_{s+1}(v)$, {{
\begin{align}
|\inr{\xi,\Gamma^*_{v} x}- \inr{\xi,\Gamma^*_{\pi_s v} \tilde{\pi}_s x}| \leq & cL \bigl( 2^{s/2}\|\Gamma^*_{v-\pi_s v} x\|_2 + 2^{s/2} \|\Gamma^*_{\pi_s v} (x-\tilde{\pi}_s x)\|_2 \bigr) \nonumber
\\
= & c L \bigl(2^{s/2}\cdot \sqrt{n} \|W(v-\pi_s v)\|_\infty + 2^{s/2}\|x-\tilde{\pi}_s x\|_{\pi_s v} \bigr) \nonumber
\\
\leq & cL \bigl(2^{s/2} \nu_s + 2^{s/2}\sup_{z \in V_r}\eps_s(z)\bigr).
\label{eq:chaining-step-2}
\end{align}}}
Iterating \eqref{eq:chaining-step-2}, summing for $s_0 \leq s < s_1$ and recalling that $2^{s_0}=\kappa_4 n/r$, it follows that with probability at least $1-2^{-c_1(L,\beta)n/r}$, 
for every $v \in V_{r,s_1}$ and every $x \in V_{k}$,
\begin{equation} \label{eq:chaining-step-2-sum}
|\inr{\xi,\Gamma^*_{v} \pi_{s_1}x -\Gamma^*_{\pi_{s_0}v} \tilde{\pi}_{s_0}x)}| \leq c_2(L,\beta)
\Bigl( \sum_{s=s_0}^{s_1-1} 2^{s/2}\nu_s + \sum_{s=s_0}^{s_1-1} 2^{s/2} \sup_{z \in V_r} \eps_s(z) \Bigr).
\end{equation}

The first sum in \eqref{eq:chaining-step-2-sum} has been estimated earlier, in \eqref{eq:gamma-wrt-infty}. In particular,
$$
c_2 \sum_{s=s_0}^{s_1-1} 2^{s/2} \nu_s \leq c_3 \sqrt{r} \alpha_r \cdot \log(c_3 r) \leq \frac{\kappa_1}{32}\sqrt{n}
$$
for $c_3=c_3(L,\beta)$ and as long as $r \leq  c_4(L,\beta)n/\log^4n$.

In order to bound the second sum in \eqref{eq:chaining-step-2-sum} we require another covering estimate.
\begin{thm} \label{thm:strong-Sudakov}
There exists an absolute constant $c$ for which, for every $z \in S^{n-1}$,
$$
\log N(V_k, \| \ \|_z , \eps) \leq c\beta\eps^{-2} k \log(en/k).
$$
\end{thm}

\proof
The proof of Theorem \ref{thm:strong-Sudakov} is an outcome of Sudakov's inequality. Fix $z \in S^{n-1}$ and define {a} linear operator $S:\R^n \to \R^n$ by $Se_i = U^i$. 
Observe that for $i \in [n]$ and $t \in \R^n$
\begin{align*}
\inr{\sqrt{n}D_{Wz}S^*t,e_i} & = \sqrt{n} \inr{S^*t,D_{Wz}e_i} = \sqrt{n} \inr{W_i,z} \inr{S^*t,e_i}\\
& = \sqrt{n} \inr{W_i,z} \inr{U^i,t}.
\end{align*}
Therefore, $\|\sqrt{n}D_{Wz}S^*t\|_2 = \|t\|_z$
and $$\log N (V_k, \| \ \|_z , \eps) = \log N (\sqrt{n}D_{Wz}S^*V_k, \eps B_2^n).$$

Set $T=\sqrt{n}D_{Wz}S^*V_k$ and let $G$ be {a} standard Gaussian vector in $\R^n$. By Sudakov's inequality (see, e.g., \cite{leta91}), there is an absolute constant $c$ for which
\begin{equation}\label{eq:Sudakov}
c \eps^2 \log N (T, \eps B_2^n) \leq \E \sup_{t \in T} \inr{G,t},
\end{equation}
{and}
\begin{align*}
\E \sup_{t \in T} \inr{G,t} &= \E \sup_{v \in V_k} \inr{G,\sqrt{n}D_{Wz}S^*v} \nonumber \\
&= \sqrt{n} \E\sup_{v \in V_k} \sum_{i=1}^n g_i \inr{W_i,z}\inr{U^i,v} \equiv \E \Bigl(\sum_{j \leq k} (Z_j^*)^2 \Bigr)^{1/2},
\end{align*}
for $Z_j =\sqrt{n} \sum_{i=1}^n g_i \inr{W_i,z}\inr{U^i,e_j}$.
Each one of the $Z_j$'s is a Gaussian variable, and since $U$ is a Hadamard type matrix with constant $\beta$,
$$
\E Z_j^2 = n \sum_{i=1}^n \inr{W_i,z}^2 \inr{U^i,e_j}^2 \leq \beta^2.
$$
Finally, {Lemma \ref{lemma:Klartag} yields
$
\E \Bigl(\sum_{j=1}^k (Z_j^*)^2 \Bigr)^{1/2} \leq c\beta \sqrt{k\log(en/k)},
$
which completes the proof by \eqref{eq:Sudakov}.}
\endproof

Invoking Theorem \ref{thm:strong-Sudakov}, {it follows that}
$$
\sup_{z \in V_r} \eps_s(z) \leq {c} \sqrt{\beta} 2^{-s/2} \sqrt{k\log(en/k)},
$$
and {due to the definition of $\alpha_r$ in \eqref{eq:alpha}} the second sum in \eqref{eq:chaining-step-2-sum} is bounded by
\begin{align*}
{\sum_{s=s_0}^{s_1} 2^{s/2} \sup_{z \in V_r} \epsilon_s(z)} 
&\leq c(L,\beta) (s_1-s_0) \sqrt{k\log(en/k)} \\
 & \leq c_1(L,\beta) \alpha_r \sqrt{k\log(en/k)} \leq \frac{\kappa_1}{32} \sqrt{n}
\end{align*}
provided that
\begin{equation} \label{eq:condition-on-k}
k \leq c_2(L,\beta) \frac{n}{\alpha_r^2 \log(e \alpha_r)}.
\end{equation}
This concludes the required estimate on $H_2$ and leads to the condition on $k$. 

The final and easiest component is to control $H_3$ in \eqref{H:decomp}. Indeed,
$$
\|\Gamma^*_{\pi_{s_0}v} \pi_{s_0}x\|_2 \leq \sqrt{n} \|W_{\pi_{s_0}}v\|_\infty \leq \beta \sqrt{r},
$$
and there are at most $2^{2^{s_0}} \cdot 2^{2^{s_0}} = 2^{2^{s_0+1}}$ pairs $(\pi_{s_0}v, \pi_{s_0}x)$.
Since $\xi$ is isotropic and $L$-subgaussian, one has that with probability at least $1-2\exp(-u^2/2)$,
$$
|\inr{\xi,\Gamma^*_{\pi_{s_0}v} \pi_{s_0}x}| \leq Lu \|\Gamma^*_{\pi_{s_0}v} \pi_{s_0}x\|_2 \leq Lu \beta \sqrt{r}.
$$
{For} $u = 2^{(s_0+2)/2}$, {the union bound yields that}, with probability at least $1-2\exp(-c2^{s_0})$,
$$
\sup_{v \in V_{r,s_1}, x \in V_k} |\inr{\xi,\Gamma^*_{\pi_{s_0}v} \pi_{s_0}x}| \leq 2^{(s_0+2)/2} L \beta \sqrt{r} \leq  \frac{2\kappa_1}{{16} \kappa_3 L \beta} \sqrt{\frac{n}{r}} \cdot L \beta \sqrt{r} \leq \frac{\kappa_1}{{8}}\sqrt{n},
$$
because {$\kappa_3 \geq 1$}. Taking the union bound over the events in which the bounds for $H_1$, $H_2$ and $H_3$ apply, we deduce that 
\[
\sup_{v \in V_{r,s_1}} \|\Gamma_v \xi\|_{[k]} \leq {\frac{\kappa_1}{4}} \sqrt{n}
\]
with probability at least $1 - 2^{-c_1 n/r} - 2 e^{-c_2 2^{s_0}} - 2^{-2^{s_1+1}} \geq 1 - 2^{-c' 2^{s_0}}$ for a suitable constant $c'$, under
condition \eqref{eq:condition-on-k} on $k$. This completes the proof of Lemma~\ref{lem:high-sparsity} after relabelling constants.
\endproof

\section{The upper bound on one-sparse vectors}
\label{sec:onesparse}

To complete the proof of Theorem \ref{thm:iid-subgaussian} one has to show that with high probability,
$$
\max_{i \in [n]} \|P_\Omega \Gamma_{e_i} \xi \|_{2} \leq c\sqrt{\delta n},
$$
for a suitable constant $c$ and $\Omega =\{i : \delta_i=1\}$, see also Theorem~\ref{thm:LM14}.

\begin{thm} \label{thm:proj-gamma-e-i}
There exist constants $c_0,c_1$ and $c_2$ that depend only on $L$ and $\beta$ for which the following holds. If $\delta \geq c_0\frac{\log n}{n}$, then with probability at least $1-2\exp(-c_1\delta n)$,
$$
\max_{i \in [n]} \left(\sum_{j=1}^n \delta_j\inr{\Gamma_{e_i}\xi,e_j}^2\right)^{1/2} \leq c_2 \sqrt{\delta n}.
$$
\end{thm}

The proof of Theorem \ref{thm:proj-gamma-e-i} is based on the fact that for every $i \in [n]$, $\Gamma_{e_i} \xi$ has a regular coordinate structure, 
in the sense that is clarified in the following lemma.
\begin{Lemma} \label{lemma:xi-mon-selectors}
There exist an absolute constant $c_1$ and a constant $c_2$ that depends on $\beta$ and $L$ for which the following holds. Let $\xi$ be an isotropic, $L$-subgaussian random vector 
and let $m \in [n]$. Then for every $i \in [n]$, with probability at least $1-\exp(-c_1m\log(en/m))$, $\Gamma_{e_i} \xi=x+y$, where
$$
\|x\|_2 \leq c_2 \sqrt{m\log(en/m)} \quad \mbox{\rm and} \quad  \max_{i \in [n]} \frac{y_i^*}{\sqrt{\log(en/i)} } \leq c_2.
$$
\end{Lemma}

\proof Let $i \in [n]$ and set $z=\Gamma_{e_i}\xi$. Let $I$ be the (random) set of the $m$ largest coordinates of $z$ and define
\begin{equation}\label{def:xz-onesparse}
x = \sum_{j \in I} z_j e_j \ \ {\rm and} \ \ y=\sum_{j \in I^c} z_j e_j.
\end{equation}
Observe that $\|x\|_2 = \|\Gamma_{e_i} \xi\|_{[m]}$. By \eqref{eq:basic-estimate-Gamma-v-2} for $p=m\log(en/m)$, and noting that $\|W e_i\|_{\infty} \leq \beta/\sqrt{n}$, one has
$$
Pr(\|x\|_{2} \geq c(L,\beta)\sqrt{m\log(en/m)}) \leq \exp(-m\log(en/m)).
$$
Repeating this argument for $m \leq \ell \leq n$, it follows that with probability at least 
$$
1-\sum_{\ell=m}^n \exp(-\ell \log(en/\ell)) \geq 1-\exp(-c_1 m \log(en/m)),
$$
for every $m \leq \ell \leq n$,
$$
z_\ell^* \leq \frac{1}{\sqrt{\ell}} \|z\|_{[\ell]} \leq 2c(\beta,L)\sqrt{\log(en/\ell)},
$$
and the claim follows.
\endproof

The coordinate structure of $\Gamma_{e_i}\xi$ comes into play thanks to a fact from \cite{meve04}.
\begin{Lemma} \label{lemma:MV-selectors}
Let $a \in \R^n$, set $\|a\|_{\psi_1^n} = \max_{1 \leq j \leq n} a_j^* /\log(en/j)$
and put $0<t<\|a\|_{\psi_1^n}/2$. Then
$$
Pr \Bigl(\bigl|\sum_{j=1}^n (\delta_j-\delta)a_j \bigr| > t \delta n \Bigr) \leq 2\exp(-ct^2 \delta n/\|a\|_{\psi_1^n}^2),
$$
where $c$ is an absolute constant and $(\delta_i)_{j=1}^n$ are independent selectors with mean $\delta$.

In particular, if $a_j=\log(en/j)$ then with probability at least $1-2\exp(-c_1\delta n)$,
$$
\sum_{i=1}^n \delta_j \log(en/j) \leq 5\delta n.
$$
\end{Lemma}

\noindent{\bf Proof of Theorem \ref{thm:proj-gamma-e-i}.}
Let $m=\delta n/\log(e/\delta)$ and consider the decomposition of $\Gamma_{e_i} \xi=x+y$ established in Lemma \ref{lemma:xi-mon-selectors}. Conditioned on the event from that lemma which holds with probability at least $1-2\exp(-c_0\delta n)$,
$$
\Bigl(\sum_{j=1}^n \delta_j x_j^2\Bigr)^{1/2} \leq \Bigl(\sum_{j=1}^n x_j^2\Bigr)^{1/2} \leq c_1(L,\beta) \sqrt{m\log(en/m)} \leq c_2(L,\beta) \sqrt{\delta n}.
$$
Also, by Lemma \ref{lemma:MV-selectors}, with probability at least $1-2\exp(-c_3\delta n)$,
$$
\sum_{j=1}^n \delta_j y_j^2 \leq c_4(L,\beta)\sum_{i=1}^n \delta_j \log(en/j) \leq c_5(L,\beta)\delta n.
$$
Hence, with probability at least $1-2\exp(-c_6\delta n)$ with respect to both $\xi$ and $(\delta_j)_{j=1}^n$, one has that
$$
\Bigl(\sum_{j=1}^n \delta_j \inr{\Gamma_{e_i}\xi,e_j}^2\Bigr)^{1/2} \leq c_7(L,\beta)\sqrt{\delta n}.
$$
Recalling that $\delta n \geq c_8\log n$ for a well chosen $c_8$, it follows from the union bound that with probability at least $1-2n\exp(-c_6\delta n) \geq 1-2\exp(-c_9\delta n)$,
$$
\max_{i \in [n]} \Bigl(\sum_{j=1}^n \delta_j \inr{\Gamma_{e_i}\xi,e_j}^2\Bigr)^{1/2} \leq c_7(L,\beta) \sqrt{\delta n}.
$$
\endproof

\section*{Acknowledgement}
The authors would like to thank Grigoris Paouris for very valuable discussions on the subject of this article. The authors would also like to thank Sjoerd Dirksen for pointing out a substantial simplification for the proof of a part of our main result, as well as the anonymous referees for their careful reading and helpful comments which greatly improved this paper.
HR was partially funded by the German Israel Foundation (GIF) through the project \textit{Analysis of structured random measurements in recovery problems} (G-1266-304.6/2014).
RW was partially funded by NSF CAREER grant \#1255631.
The authors acknowledge support during their stay at the trimester program, "Mathematics of Signal Processing" at the Hausdorff Research Institute for Mathematics, University of Bonn in 2016, where parts of this work have been developed.

\bibliographystyle{abbrv}
\bibliography{MPRW}

\end{document}